\documentclass[12pt,preprint]{aastex}
\newcommand{\emp}{\rm}

\newcommand{\be}{\begin{equation}}
\newcommand{\ee}{\end{equation}}
\newcommand{\bea}{\begin{eqnarray}}
\newcommand{\eea}{\end{eqnarray}}

\newcommand{\di}{\partial_}

\shorttitle{Relaxed Dark Matter Halos }
\shortauthors{Henriksen}

\begin{document}

\title{Central Relaxation of  Dark Matter Halos}

\author{R. N. Henriksen}
\affil{Department of Physics, Engineering Physics and Astronomy,  Queen's University, Kingston, Ontario,Canada, K7L 3N6}
\email{henriksn@astro.queensu.ca}

\begin{abstract}
In this paper we  reconsider a series expansion for a dark matter distribution function in the spherically symmetric anisotropic limit. We show here  that the expansion may be renormalized so that the series does converge in time to an estimate of the steady state distribution function in the central regions. Subsequently  we use this distribution function to discuss the nature of the central equilibrium and, by invoking stationarity of Boltzmann's H function as a measure of (thermodynamic) relaxation, we calculate the adiabatic variation in the  local logarithmic slope of the mass density. Similarly the pseudo (phase-space) density variation with radius is calculated.  These are compared to  empirical fitting functions. There is general agreement on the inner part of the logarithmic slope of the density and of the inner profile of the pseudo-density power law, but coincident continuity with the outer power-laws is not yet achieved.    Finally some suggestions are made regarding the actual microphysics acting during the non-equilibrium approach to relaxation. In particular a cascade regime is identified.     
\end{abstract}



\section{INTRODUCTION}

This paper continues the study of the dynamical relaxation of collisionless matter using the analytical technique summarized recently in Henriksen (2006b: paper I hereafter). The reader is encouraged to refer to paper I for a more detailed introduction to the method, but we repeat the general idea here for convenience.

 We transform from the usual phase-space variables plus time to what might be called `Lie' variables in that they remain invariant under a Lie displacement (see Carter and Henriksen, 1991 for a formal treatment; or Henriksen, 1997 for pragmatic application to the present problem). The direction of this displacement in phase-space-time is taken to be along one of the variables ($T$ in the present paper) and {\it the symmetry is only imposed by insisting that} $\partial_T=0$. Otherwise the transformation is merely a  transformation to rotated variables in phase-space-time with no loss of generality.    

In the application to the dark matter halo problem, it is convenient to remain close to the homothetic or self-similar symmetry . This is a Lie symmetry wherein the relation between the physical quantities and  the invariant quantities is found essentially by dimensional analysis. Because of the inevitable presence of Newton's constant $G$, the complete set of dimensions is reduced essentially to space and time. These are scaled independently by the parameters $\delta$ and $\alpha$ respectively, but all physical results depend only on the ratio $a\equiv \alpha/\delta$. 

More explicitly,Lie scaling in the  dimension space of time, space and mass is represented by the scale vector $(\alpha,\delta,\mu)$. This can be reduced to the scale vector ${\bf A}\equiv(\alpha,\delta)$ since the mass scale-factor $\mu\equiv (3\delta-2\alpha)$ due to the invariance of Newton's constant, $G$. The usual dimensional analysis of each dynamical quantity $q$ is effected by assigning the appropriate dimensional co-vector ${\bf d}_q$ to $q$.
Thus for a velocity ${\bf d}_v=(-1,1)$, while for density ${\bf d}_\mu=(-2,0)$ after eliminating $\mu$ in favour of $3\delta-2\alpha$.    

 The independent variables that express this transformation are 
\bea
R\equiv r~ e^{-(\alpha T)/a},& & Y\equiv v_r/(\alpha t)^{(1/a-1)}\nonumber\\
Z\equiv j^2/(\alpha t)^{(4/a-2)},& &\alpha T\equiv \ln{\alpha t},\\ \label{vars}
\nonumber\eea
where the usual notation for radius, radial velocity, squared specific angular momentum, time, and the distribution function are used. These variables are defined to be independent of $T$. The variation of any quantity $q$ under the scaling transformation is given by $\partial_T q=({\bf d}_q.{\bf A})q$, which integrates to $q=q(R,Y,Z;T)\exp{({\bf d}_q.{\bf A}~T)}$. Thus taking $q=G$ and observing that ${\bf d}_G=(-2,3,1)$, we deduce that $3\delta-2\alpha=\mu$ for the invariance of $G$ under the scaling as used above.   

 The strict self-similarity (now identified as a Lie symmetry) is given by holding $a$ constant and setting $\partial_T=0$. The advantage of these variables is therefore clear in this limit since the problem is now rendered `stationary' although it remains multi-dimensional.  By remaining `close' to the self-similar symmetry we mean that there is only an adiabatic variation in $a$ locally due to dynamical modification of the initial conditions. We recall that for a strict self-similarity set by the cosmological boundary conditions $P(k)\propto k^n$, where $n$ is appropriate to a given range of $k$, one finds (e.g. Henriksen and Widrow, 1999; Hoffman and Shaham, 1985) 
\be
a=\frac{3(3+n)}{2(x+n)},\label{asim}
\ee
where $x=4$ around a local maximum (Hoffman and Shaham, 1985) or $x=5$ around an $n\sigma$ peak (Henriksen,2006b).  

Except in situations of extreme phase-space symmetry this formulation does not lead to exact solutions of the Poisson-Boltzmann set of equations even when $\partial_T=0$ . To proceed further analytically we make use of the freedom of either one of the parameters $\alpha$ or $\delta$, here taken to be $\alpha$, to regulate the resolution in phase-space and in time according to the relation for the phase-space volume
\be
\Delta r~\Delta v_r~\Delta j^2=\Delta R~\Delta Y~\Delta Z~e^{(6/a-3)\alpha T},
\ee
and the to definition of $T$ in terms of $t$. By allowing $\alpha$ to be large we  coarse-grain the phase-space resolution as well as degrade the temporal resolution (an averaging). Thus this transformation is necessarily non-canonical from which fact it derives its usefulness.

We thereby obtain a convenient and well-defined coarse-graining procedure, which we apply  by writing all quantities as a series in inverse powers of $\alpha$ and solving progressively for the various orders. In the past we have  terminated the series at a given order by requiring all higher order terms to vanish. However the approximation found in this way diverges in time at each radius for any finite $\alpha$. Thus our first task in this paper is to find a suitable renormalization procedure that renders high order terms finite asymptotically in time at finite $\alpha$. 

 In paper I the distribution function of dark matter was studied in just this manner.  
The series expansion for the distribution function $f$ takes the explicit form to second order (see equation (18) of paper I)

\be
\pi f(r,v_r,j^2;t)=r^{(a-3)}(P_{oo}(\zeta_1,\zeta_2^2;T)-u~P_{11}(\zeta_1,\zeta_2^2;T)+(u^2/2)~P_{22}(\zeta_1,\zeta_2^2;T),\label{series}
\ee
where $u$ is defined as 
\be
u\equiv 1/(\alpha R^a)=t/r^a, \label{u}
\ee
 and the factor $\pi$ is for convenience only. We use the variable $u$ in the renormalization procedure below.

This expression uses the dimensional scaling 
\be
\pi f=P(R,Y,Z)e^{-(3/a-1)\alpha T},\label{varf}
\ee 
where 
\be
P=P_o+P_1/\alpha+P_2/\alpha^2+\dots,
\ee
so that 
\bea
P_o\equiv P_{oo}r^{(a-3)},&~&P_1\equiv- P_{11}r^{(a-3)}/R^a,\nonumber\\
                          &~&P_2\equiv P_{22}r^{(a-3)}/(2R^{2a}).\label{Pjj}
\eea

The functions $P_{jj}$ are functions of the new variables (\ref{vars})
in the combinations 
\be
\zeta_1\equiv YR^{(a-1)},~~~~~~~~~~\zeta_2^2\equiv ZR^{2(a-2)}.\label{charvars}\ee 
These latter variables are constants on the characteristics of the collisionless Boltzmann equation (CBE) to all orders. 

When approximate or `adiabatic' self-similarity is assumed, the  dependence on $T$ indicated in the functions $P_{jj}$ is weak (absent for strict self-similarity). However this dependence will be allowed in the renormalization procedure.
The parameter $a$ is the self-similarity index. In the case of self-similarity determined by initial conditions it gives the ratio of the temporal to the spatial powers in the dimensions of a governing constant. However as argued in paper I and in Henriksen (2006a), this dominant constant can change dynamically. It is then best left as a fitting parameter. We discuss its variation further in susequent sections.

The functions $P_{oo}$, $P_{11}$ and $P_{22}$ are related from the expansion of the Poisson and Boltzmann equations according to

\be
P_{11}\equiv \left((a-1)(\zeta_1^2-2\gamma_o)+\zeta_2^2\right)\di {\zeta_1} P_{oo}+2(a-2)\zeta_1\zeta_2^2\di {\zeta_2^2}P_{oo}-\zeta_1(3-a)P_{oo},
\label{P11}
\ee 

and 
\be
P_{22}\equiv ((a-1)(\zeta_1^2-2\gamma_o)+\zeta_2^2)\di {\zeta_1}P_{11}+2(a-2)\zeta_1\zeta_2^2\di {\zeta_2^2}P_{11}-3\zeta_1P_{11}+(3a-2)\gamma_1\di {\zeta_1}P_{oo}.
\label{P22}
\ee

The constants occurring in the potential, $\gamma_o$ and $\gamma_1$, are defined as in paper I by 
\be
\gamma_o\equiv \frac{I_{oo}}{2(a-1)(3-2a)},\label{gammao}
\ee
and 
\be
\gamma_1\equiv \frac{I_{11}}{3(1-a)(2-3a)},\label{gamma1}
\ee
where 
\be
I_{jj}\equiv \int~P_{jj}~d\zeta_1 d\zeta_2^2.
\ee

The potential is given to first order by 
\be
\Phi=-r^{2(1-a)}(\gamma_o+\gamma_1 u+---),\label{pot}
\ee
where the general dimensional scaling between $\Phi$ and the `invariant' potential $\Psi$ is 
\be
\Phi(r,t)\equiv \Psi(R,T)e^{2(1/a-1)}
\ee
The function $\Psi$ has been expanded as $\Psi=\Psi_o+\Psi_1/\alpha+\dots$ and $\Psi_o$ and $\Psi_1$ can be identified from equation (\ref{pot}). 

The  series corresponding to equation (\ref{series}) for the density is (to second order)
\be
\rho=r^{-2a}(I_{oo}-I_{11}u+\frac{I_{22}}{2}u^2---),\label{density}
\ee
where the relation between $\rho$ and the invariant density $\Theta$ is  
\be
\rho =\Theta e^{-2\alpha T},
\ee
and $\Theta$ has also been expanded as $\Theta_o+\Theta_1/\alpha+\dots$.

 One notes however that in the context of adiabatic self-similarity where $a=a(r)$ the dependence of the density on $r$ enters  through the dependence of the $I_{ij}$ on $a$ as well as through $a(r)$ in the exponent.

Normally we terminate the series by insisting that $P_{22}=0$ (paper I) whereupon equations (\ref{P11}) and (\ref{P22}) become coupled partial equations for $P_{oo}$ and $P_{11}$. However we proceed differently here.

The difficulty with the series expansion (\ref{series}) lies in the time dependence of $u$. This quantity increases monotonically in time and so each of the terms in the series eventually becomes too large to be neglected. Hence one can only expect transitory validity to such an expansion. 

It is important to do better than this since this series was used to discuss the power-law behaviour of the phase space pseudo-density (paper I). It so happens that recently McDonald (2006, 2007) addressed this problem in the context of the cosmological perturbative approach to dark matter clustering. This technique replaces the different transitory  estimates of the unknown function by an envelope to all such transitory functions. Unlike the individual transitory functions, the envelope is valid asymptotically. This technique is very well described in the paper by Kunihiro (1995), and we adapt this method to our problem in the next section. This renormalization procedure is somewhat technical and does not change our previous conclusions. The reader may wish to skip to section (4.1) where the renormalized distribution function is given together with a brief description of the method. The last section of the paper together with the discussion introduce our new physical ideas and give numerical results.

\section{Geometrical Renormalization}

  We take the approximate functions of $u$ for which we seek the envelope to be ($u_o$ is a parameter) to zeroth order 
\be
r^{(3-a)}F\equiv P_{oo}-P_{11}(u-u_o),\label{1stenvelope}
\ee
and to first order

\be
 \tilde F\equiv \frac{(F-r^{(a-3)}P_{oo})}{(ur^{(a-3)})}=-P_{11}+\frac{P_{22}}{2}(u-u_o),\label{envelope}  
\ee
where $F\equiv \pi f$.

The procedure that yields the envelope function (Kunihiro, 1995) consists in differentiating the right-hand side of this equation with respect to $u_o$ and subsequently setting $u_o=u$ in order to find the envelope at $u$. This yields evidently to zeroth order

\be
\di{u}P_{oo}+P_{11}=0,\label{renormequ1}
\ee
and to first order
\be
\di{u}P_{11}+P_{22}/2=0.\label{renormequ2}
\ee

These last two equations replace the previous termination that use $P_{11}=0$ and $P_{22}=0$ to zeroth and first order respectively.  Together with equation (\ref{P11}) to zeroth order plus equation (\ref{P22}) to first order we obtain a partial differential set for $P_{11}(u,\zeta_1,\zeta_2^2)$, $P_{oo}(u,\zeta_1,\zeta_2^2)$ and finally $P_{22}(u,\zeta_1,\zeta_2^2)$ if carried out to first order. One seeks to choose the arbitrary functions appearing in the solutions to these equations so that the expression for $F$ becomes asymptotically independent of $u$ after setting $u=u_o$. This latter identity is the reason we obtain $F$ only to zeroth and first order even though we retain expansion terms up to first and second order respectively.

Equation (\ref{renormequ1}) together with  equation (\ref{P11}) yields  the relevant  equation for the $P_{oo}$ and $P_{11}$ which we write here for explicitness as:
\be
 \left((a-1)(\zeta_1^2-2\gamma_o)+\zeta_2^2\right)\di {\zeta_1} P_{oo}+2(a-2)\zeta_1\zeta_2^2\di {\zeta_2^2}P_{oo}-\zeta_1(3-a)P_{oo}+\di{u}P_{oo}=0.\label{E1}
\ee
This procedure will give the original series for the DF to zeroth order in a renormalized form as discussed in the next section.

If we wish to continue to first order renormalization, we must replace the final term on the left in equation (\ref{E1}) by $-P_{11}$ to obtain
\be
\left((a-1)(\zeta_1^2-2\gamma_o)+\zeta_2^2\right)\di {\zeta_1} P_{oo}+2(a-2)\zeta_1\zeta_2^2\di {\zeta_2^2}P_{oo}-\zeta_1(3-a)P_{oo}-P_{11}=0.\label{E11}
\ee
 and then add equation (\ref{renormequ2}) combined with equation (\ref{P22}) in the form
\be
2\di{u}P_{11}+((a-1)(\zeta_1^2-2\gamma_o)+\zeta_2^2)\di {\zeta_1}P_{11}+2(a-2)\zeta_1\zeta_2^2\di {\zeta_2^2}P_{11}-3\zeta_1P_{11}+(3a-2)\gamma_1\di {\zeta_1}P_{oo}=0.   \label{E2}
\ee
After equations (\ref{E11}) and (\ref{E2}) are solved for $P_{oo}$ and $P_{11}$, equation (\ref{renormequ2}) yields $P_{22}$. However this latter term will not appear in the DF expression after setting $u=u_o$.
 
In practice, we normally use the zeroth order solution from equation (\ref{E1}) in equation (\ref{E2}) to start an iteration process that yields a non-zero $P_{11}$. The process is continued by using in equation (\ref{E11}) the $P_{11}$ found in this way  to obtain a new $P_{oo}$.   
We shall discuss these renormalizations in turn in the following sections. 

\section{Zeroth order renormalization}

We solve equation (\ref{E1}) by the method of characteristics, so that we must solve the set
\bea
\frac{d\zeta_1}{du}&=& (a-1)(\zeta_1^2-2\gamma_o)+\zeta_2^2,\nonumber\\
\frac{d\zeta_2^2}{du}&=& 2(a-2)\zeta_1\zeta_2^2,\label{chars}\\
\frac{dP_{oo}}{du}&=&(3-a)\zeta_1P_{oo}.\nonumber\\
\nonumber\eea

These integrate to give
\be
P_{oo}=K(\kappa_1,\kappa_2){\cal E}_o^{\left(\frac{3-a}{2(a-1)}\right)},\label{RePoo}
\ee
where 
\be
\kappa_1\equiv {\cal E}_o(\zeta_2^2)^{-(\frac{1-a}{2-a})},\label{kappa1}
\ee
and $\kappa_2$ must be found from
 
\be
\int~\frac{d\zeta_2^2}{\zeta_2^2(\sqrt{2\kappa_1(\zeta_2^2)^{(\frac{1-a}{2-a})}-\zeta_2^2-2|\gamma_o|})}\pm 2(2-a)u\equiv \kappa_2.\label{kappa2}
\ee

The function $K(\kappa_1,\kappa_2)$ is arbitrary and 
\be
{\cal E}_o\equiv |\frac{\zeta_1^2+\zeta_2^2}{2}-\gamma_o|.\label{Eo}
\ee

The integral in equation (\ref{kappa2}) is not in general expressible in simple terms (although it is in the special case $a=0$). However  the entire interesting range of $a$ is $[0,3/2]$ (the outer density logarithmic slope being $-3$)  and in this paper, in order to describe a flattened density profile (relative to $-2$), we may consider here $a< 1$. Then in this domain we observe that the integral can be expected to be periodic in $\zeta_2$ since the square root will impose inner and outer `turning' points. The quantity $\kappa_2$ is not then an isolating integral, but gives rather the temporal behaviour of the (scaled)` angular momentum' on the characteristics, which are not in these variables  the particle trajectories. For our purposes the important behaviour is simply the monotonic linear dependence on $u$.

If at this order we wish to have an asymptotic approach to the  steady state, then remembering equation (\ref{RePoo}) we may require for example that
\be
K(\kappa_1,\kappa_2)=K_1(\kappa_1)K_2(\frac{\kappa_2}{constant+\kappa_2}).\label{Ko}
\ee
But any form of $K_2$ and its argument that tends to a constant asymptotically will do. If in addition we require a steady, spherically symmetric and velocity isotropic asymptotic DF, we should set $K_1$ $=$ constant . 
Hence the zeroth order, that is steady state, distribution function becomes  
\be
F=r^{(a-3)}K{\cal E}_o^{(\frac{(3-a)}{2(a-1)})},
\ee
which is explicitly
\be 
\pi f=C_{oo}|\frac{v_r^2}{2}+\frac{j^2}{2r^2}+\Phi_o(r)|^{\frac{(3-a)}{2(a-1)}},
\label{zerothDF}
\ee
as in paper I with $C_{oo}$ constant. This is thus our best estimate for the ultimate DF in an isotropic, spherically symmetric, adiabatically self-similar (i.e. $a$ varies slowly in a crossing time) dark-matter core. 

We do not thus change our  conclusions in paper I, but we obtain some assurance that the series can asymptote to the steady value.
\section{First order renormalization}

If we wish to find a stable approximation to the approach to the steady state, then we must renormalize the first order as well as the zeroth order simultaneously.  This requires solving equation (\ref{E2}) using the zeroth order renormalized $P_{oo}$. Hence we solve the set of characteristic equations (see also paper I);
\bea
\frac{d\zeta_1}{ds}&=& (a-1)(\zeta_1^2-2\gamma_o)+\zeta_2^2,\nonumber\\
\frac{d\zeta_2^2}{ds}&=& 2(a-2)\zeta_1\zeta_2^2,\label{chars1}\\
\frac{dP_{11}}{ds}&=&3\zeta_1P_{11}-(3a-2)\gamma_1(\di{\zeta_1}P_{oo})_{char},\nonumber\\
\nonumber\eea
where $P_{oo}$ is taken from the zeroth order renormalization (\ref{RePoo}) and $ds\equiv du/2$.

As in the previous section these equations yield the characteristic constants $\kappa_1$ and $\kappa_2$ (although with $u$ in equation (\ref{kappa2}) replaced by $s$). The equation for $P_{11}$ on the characteristics becomes, by using the third member of equations (\ref{chars1}), equation (\ref{RePoo}), and taking $\zeta_2^2$ as the variable, 
\bea
&\zeta_2^2&\frac{dP_{11}}{d\zeta_2^2}= \label{P11equ}\\
&-&\frac{3}{2(2-a)}P_{11}+\frac{(3a-2)}{2(2-a)}\gamma_1K(\kappa_1,\kappa_2){\cal E}_o^{(p-1)}\left(p+\di{\ln{\kappa_1}}\ln{(K)}-\kappa_1\di{\kappa_2}\ln{(K)}g(\zeta_2^2)\right),\nonumber\\
\nonumber\eea
where
\be
p\equiv \frac{(3-a)}{2(a-1)},\label{p}
\ee
and 
\be
g(\zeta_2^2)\equiv \int^{\zeta_2^2}~dx~\frac{1}{x^{\frac{1}{2-a}}(2\kappa_1x^{\frac{1-a}{2-a}}-x-2|\gamma_o|)^{3/2}}.
\ee

It becomes clear from equations (\ref{envelope}) and (\ref{P11equ}) ( remembering the linear, monotonic dependence of $\kappa_2$ on $u$) that only the choice 
\be
K(\kappa_1,\kappa_2)=K_1(\kappa_1)/\kappa_2,\label{renormK}
\ee
will produce an envelope behaviour ($F$ independent of $u=u_o$) to first order. This also implies that the term proportional to $\di{\kappa_2}K$ in equation (\ref{P11equ}) may be dropped asymptotically at large $u$. In this case the integration for $P_{11}$ proceeds as in paper I (e.g. equation (35) in that paper) so we obtain formally (the constant of integration on the characteristics $\tilde K(\kappa_1,\kappa_2)$ is also written as $\tilde K=\tilde K_1(\kappa_1)/\kappa_2$)
\be
P_{11}=\frac{1}{\kappa_2}\left(\gamma_1K_1(\kappa_1)(p+\frac{d\ln{K_1}}{d\ln{\kappa_1}}){\cal E}_o^{\frac{5-3a}{2(a-1)}}+\tilde K_1(\kappa_1){\cal E}_o^{\frac{3}{2(a-1)}}\right).\label{ReP11}
\ee
In the special case $a=2/3$, $\gamma_1$ must be replaced formally by $-(3I_{11}/2)\ln{{\cal E}_o}$.
 In any case we recall that $\gamma_1\propto I_{11}$ and hence to  $P_{11}$  as written immediately above, which vanishes asymptotically with $1/\kappa_2$. Thus we may ultimately neglect this term compared to the second term in the bracket of equation (\ref{ReP11}). However it is useful to retain it as a transitory term (see below) in the resulting distribution function. 

Strictly asymptotically then we deduce for $P_{11}$
\be
P_{11}=\frac{1}{\kappa_2}\tilde K_1(\kappa_1){\cal E}_o^{\frac{3}{2(a-1)}}.\label{1stP11}
\ee

We turn now to the modified equation (\ref{E1}) to obtain the first order correction from 
\be
\frac{dP_{oo}}{ds}=(3-a)\zeta_1P_{oo}+P_{11},
\ee
which, with the substitution 
\be
P_{oo}=C_{oo}(s)(\zeta_2^2)^{-\frac{3-a}{2(2-a)}},\label{1stPoo}
\ee
becomes
\be
\frac{dC_{oo}}{ds}=(\zeta_2^2)^{-\frac{3-a}{2(2-a)}}P_{11}.
\ee 

To complete the explicit solution for $P_{oo}$ is evidently complicated since we require 
\be
\frac{d\zeta_2^2}{ds}=2(a-2)\zeta_1\zeta_2^2,
\ee
from equations (\ref{chars1}) and 
\be
\zeta_1^2=2\kappa_1(\zeta_2^2)^{\frac{1-a}{2-a}}-\zeta_2^2-2|\gamma_o|,
\ee
from the integral (\ref{kappa1}) and the definition (\ref{Eo}). Moreover the integral $\kappa_1$ allows us to eliminate ${\cal E}_o$ in favour of $\zeta_2^2$ from $P_{11}$. The integrals over $\zeta_2^2$ are not expressible in general in terms of elementary functions, but they are in any case periodic because of the turning points in $\zeta_1$. Ultimately they provide too much information, since $P_{11}$ vanishes asymptotically. This means that $C_{oo}$ is asymptotically constant and thus $P_{oo}$ retains the same zeroth order form as in the previous section. Consequently the distribution function renormalized to first order is 
\be
 F=r^{(a-3)}\left(C_{oo}(\kappa_1){\cal E}_o^{\frac{(3-a)}{2(a-1)}}- uP_{11}\right).\label{renormedF}
\ee
This is translated into the physical form that we use in our numerical discussions in the next sub-section.

\subsection{Summary of Renormalization Result}

The renormalization procedure followed in the previous sections due to Kunihiro (1995) comprises first finding an approximation to some function of interest (found generally by a series expansion as in our equation (\ref{series})). In general the approximation may approach the desired function only in a transitory fashion in the independent variable ($u$ in our case). Subsequently the approximation is parameterized as in our equations (\ref{1stenvelope}) and (\ref{envelope}) in such a fashion that the value of the parameter ($u_o$ in our case) determines the region over which the approximation holds. The envelope to this set of functions (created as $u_o$ varies) is then found by standard methods of the calculus plus the ingenious device of setting $u_o=u$ after differentiating with respect to this parameter.

The procedure is somewhat special in our case in that the terms in the approximation (the $P_{jj}$) are related already through equations (\ref{P11}) and (\ref{P22}), and the renormalization conditions (\ref{renormequ1}), (\ref{renormequ2}) merely close the system at the respective orders. The previous sections have thus been occupied with the solution at zeroth and first orders for $P_{oo}$ and $P_{11}$ which render the corresponding series (\ref{1stenvelope}) and (\ref{envelope}) independent of $u$.

This procedure concluded with the result of equation (\ref{renormedF}) (together with equation (\ref{ReP11}) which in terms of physical quantities becomes explicitly 
\bea
&\pi A f=\label{result1}\\
&C_{oo} A|E|^{\frac{3-a}{2(a-1)}}-\tilde K_1(\kappa_1)|E|^{\frac{3}{2(a-1)}}r^a -\gamma_1K_1(\kappa_1)\left(\frac{3-a}{2(a-1)}+\frac{d\ln{K_1}}{d\kappa_1}\right)|E|^{\frac{5-3a}{2(a-1)}}r^{(2-2a)}.\nonumber\\
\nonumber\eea
Here 
\be
|E|=|\frac{v_r^2}{2}+\frac{j^2}{2r^2}+\Phi_o(r)|,
\ee
 a $\pm$ sign has been absorbed in the constants $K$, and $A\equiv 2(2-a)$.

Once again for $a=2/3$, we must formally replace $\gamma_1$ by $-(3I_{11}/2)(\ln{|E|}+\ln{r^{-2/3}})$, although this term vanishes in time in any case as remarked previously and is strictly of transitory interest at a fixed $r$.

Equation (\ref{result1}) gives the  formal result of this paper.  Thus asymptotically, when the time dependent term disappears, the first order term does not dominate the zeroth order term as $r\rightarrow 0$. This was not the case for the original unrenormalized series (\ref{series}). There is however still a limit to the validity of the asymptotic (in time) expression at large $r$ where the first order term becomes comparable to the zeroth order term.

We have left the decreasing time dependent term in $\gamma_1$ (actually $I_{11}$) because of the  $r$ dependence at  $a<1$. This may be used to off-set the time dependence (that is we need $r^{(2-2a)}\propto u$, which  is $r\propto t^{1/(2-a)}$ to keep this term's importance relative to the zeroth order).  Thus this term allows some estimate of the approach to the asymptotic form at any fixed $r$. The dominant first order term proportional to $\tilde K_1$ is renormalized in time but as already noted gives an outer limit to the validity of the series when it becomes comparable to the zeroth order. We normally take this term to be negligible at small $r$ in what follows and give our preliminary applications uniquely in terms of the zeroth order (equilibrium) part of (\ref{result1}). The time it takes for this equilibrium to be attained may be estimated in principle at each $r$ by insisting that the transitory term be negligible, but this would require a precise determination of the constants for a particular system.   

In the next section we discuss especially the implications of the zeroth order  for the central density and pseudo-density profiles, including the adiabatic variation of $a$.

\section{Adiabatic Approach to Central Equilbrium}

Taking the zeroth order result from the previous section as the central distribution function of a dark matter halo, we calculate the density  from the integral
\be
\rho(a,r)=2\sqrt(2)C_{oo}\int_{\Phi_o}^{E_m}~dE~\sqrt{(E-\Phi_o)}~E^{\frac{3-a}{2(a-1)}},
\ee

which yields after the substitution $u\equiv \Phi_o/E$ 
\be
\rho(a,r)=2\sqrt{2}C_{oo}|\gamma_o|^{\frac{a}{a-1}}r^{-2a}~D(a;u_m),\label{rho}
\ee
where the {\it ad hoc} function $D$ is 
\be
D(a;u_m)\equiv \int_{u_m}^1~du~\sqrt{1-u}~u^{\frac{2a-1}{1-a}},
\ee
or
\be
D(a;u_m)=B(\frac{a}{1-a},3/2)-\frac{1-a}{a}~u_m^{\frac{a}{1-a}}~_2F_1(\frac{a}{1-a},-1/2;\frac{1}{1-a};u_m).\label{F}
\ee

 We define $u_m$ as 
\be
u_m\equiv \Phi_o/E_m,
\ee
where $E_m$ is the maximum energy present at $r$.
 
The quantity $|\gamma_o|$ is found by completing the integral for $I_{oo}$ as in paper I and using the definition (\ref{gammao}) to be 
\be
|\gamma_o|^{\frac{1}{1-a}}=\frac{2\sqrt{2}C_{oo}}{(1-a)(3-2a)}~D(a;u_m),
\ee
and this in turn gives  $\Phi_o(a,r;u_m)$ from equation (\ref{pot}).

In these expressions $B(x,y)$ is the Euler Beta function and $_2F_1(\alpha,\beta;\gamma;z)$ is the Gaussian hypergeometric function.

 This gives the density explicitly as a function of $a$ and $r$ in the form
\be
\rho=(kD)^{(1-a)}((1-a)(3-2a))^a~r^{-2a}.\label{Xdensity}
\ee
Hence the logarithmic slope $\beta=-dlog{\rho}/dlog{r}$ is found as
\bea
 &\beta&=2a\times \nonumber\\
(\!\!\!\!\!&1&\!\!\!\!\!+\frac{\alpha_T}{2}(\ln{(kD(a,U_m)r^2)}\!-\!\frac{d\ln{(kD(a;u_m))^{(1-a)}}}{da}-
\ln{(3\!-\!5a\!+\!3a^2)}\!+\!\frac{a(5-4a)}{(3\!-\!5a\!+\!2a^2)}))\!\!\label{betaT}
\eea
for $a<1$, where we have defined $\alpha_T\equiv d\ln{a}/d\ln{r}$ and $k\equiv 2\sqrt{2}C_{oo}$.

Clearly if $\alpha_T$ is small $\beta\approx 2a(r)$. This tends to be the case except at  small $k$, which correponds to a small central density and/or at relatively large $r$. In such cases our argument below yielding $a(r)$ in terms of the Boltzmann H function, probably fails in any case because of insufficient relaxation.

Indeed an important consideration is that only with $u_m \ne 0$ does one obtain a finite value for $\rho$ as $a$,$r$ both tend to zero. Otherwise the Beta function in $D(a,u_m)$ dominates and the density diverges as $1/a$ ($B(x,3/2)\approx 1/x$ as $x\rightarrow 0$). Although with $u_{mc}\equiv lim_{r\rightarrow 0}u_m=const.~\ne 0$ the central density is always finite, it  becomes only logarithmically larger as $u_{mc}$ approaches zero since the cancellation of the $1/a$ divergence in $D$ requires $|a\ln{u_{mc}}|\approx 1$. 

However, using $\Phi_o$ in the definition explicitly, $u_{mc}=\lim_{(r\rightarrow 0)}(|\gamma_o|r^{2(1-a)}/E_m)$ where $E_m$ is the maximum energy present at $r$.  Unless $E_m\rightarrow 0$ in the same limit $u_{mc}$ goes to zero as $r\rightarrow 0$ ($a<1$), which by the above discussion would imply an infinite central density. We therefore conclude that $E_m\propto \Phi_o$ as $r\rightarrow 0$  (so that both quantities vanish at the centre at the same rate) if one is to have a finite central density. The energy dispersion (and hence the velocity dispersion) of the central particles will thus decrease ultimately towards zero.

From the empirical fitting formula of Navarro et al. (2004) we deduce that 
\be
\beta=2(\frac{r}{r_{-2}})^{\alpha_F},\label{betaF}
\ee
where $r_{-2}$ is the radius where $\beta=-2$ and $\overline{\alpha_F}\approx 0.17$, although there is dispersion about this value at the $20\%$ level. If  one ignores the terms multiplying $\alpha_T$ in equation (\ref{betaT}) we find
 
\be
a=(\frac{r}{r_{-2}})^{\alpha_F}.\label{a}
\ee  
  Thus  $a$ and $r$ do tend to zero together (and $a$ is adiabatically slow relative to $r$) according to the numerical simulations and consequently in our expression $u_m\ne 0$, however small, in order to guarantee a finite central density. One can imagine that such delicate physical behaviour may well be reflected in the difficult convergence of numerical simulations.

Also from (\ref{result1}) we may calculate the rms isotropic velocity dispersion $\sigma^2$ from $\overline{v^2}-\overline{v}^2$, that is
\be
\sigma^2= \frac{4}{\rho} \int_{\Phi_o(r)}^{E_m}~f((2(E-\Phi_o(r)))^{3/2}-2(E-\Phi_o))~dE,
\ee
which  yields at equilibrium (zeroth order)
\be
\sigma^2=4\left(\frac{k}{(1-a)(3-2a)}\right)^{(1-a)}r^{2(1-a)}\left(\frac{G_b(a;u_m)}{D(a;u_m)^a}-\frac{2{G_1(a,u_m)}^2}{D(a;u_m)^{(1+a)}}\right).\label{rms}
\ee

Here we have defined  new functions
\be
G_b(a;u_m)\equiv (\frac{1}{q})[x^{q}~_2F_1(q,-3/2;\frac{a}{1-a},x)]_{u_m}^b,\label{G}
\ee
where $b=1_-$ and 
\be
q(a)\equiv \frac{1-a}{2a-1},
\ee

plus
\be
G_1(a;u_m)\equiv \frac{2(1-a)}{(1+a)}\left(\frac{(1+a)}{(3a-1)}(1-u_m^{\frac{(3a-1)}{2(1-a)}})-(1-u_m^{\frac{(1+a)}{2(1-a)}})\right).\label{G1}
\ee
With the adiabatic variation $a(r)$ the $r$ dependence of this expression for $\sigma^2$ is no longer obvious. Despite appearances, it is well behaved at $a=1/2$. 

The quantity $u_{mc}$ determines the rate at which the particle energy decreases to zero at the centre of the system since $\Phi_o$ vanishes there as discussed previously. 
For simplicity we have taken $u_m$ to be  constant everywhere in our numerical results rather than just at the centre of the system, but it might vary with $r$ and $a(r)$ through the dependence  
\be
u_m\equiv\Phi_o/E_m=\left(\frac{kr^2}{(1-a)(3-2a)}\right)^{(1-a)}\frac{D(a;u_m)}{E_m}\label{um}
\ee 
if $E_m$ is instead taken constant for all $r$. 

\begin{figure}[h]
\begin{tabular}{cc} 
\rotatebox{0}{\scalebox{.5} 
{\includegraphics{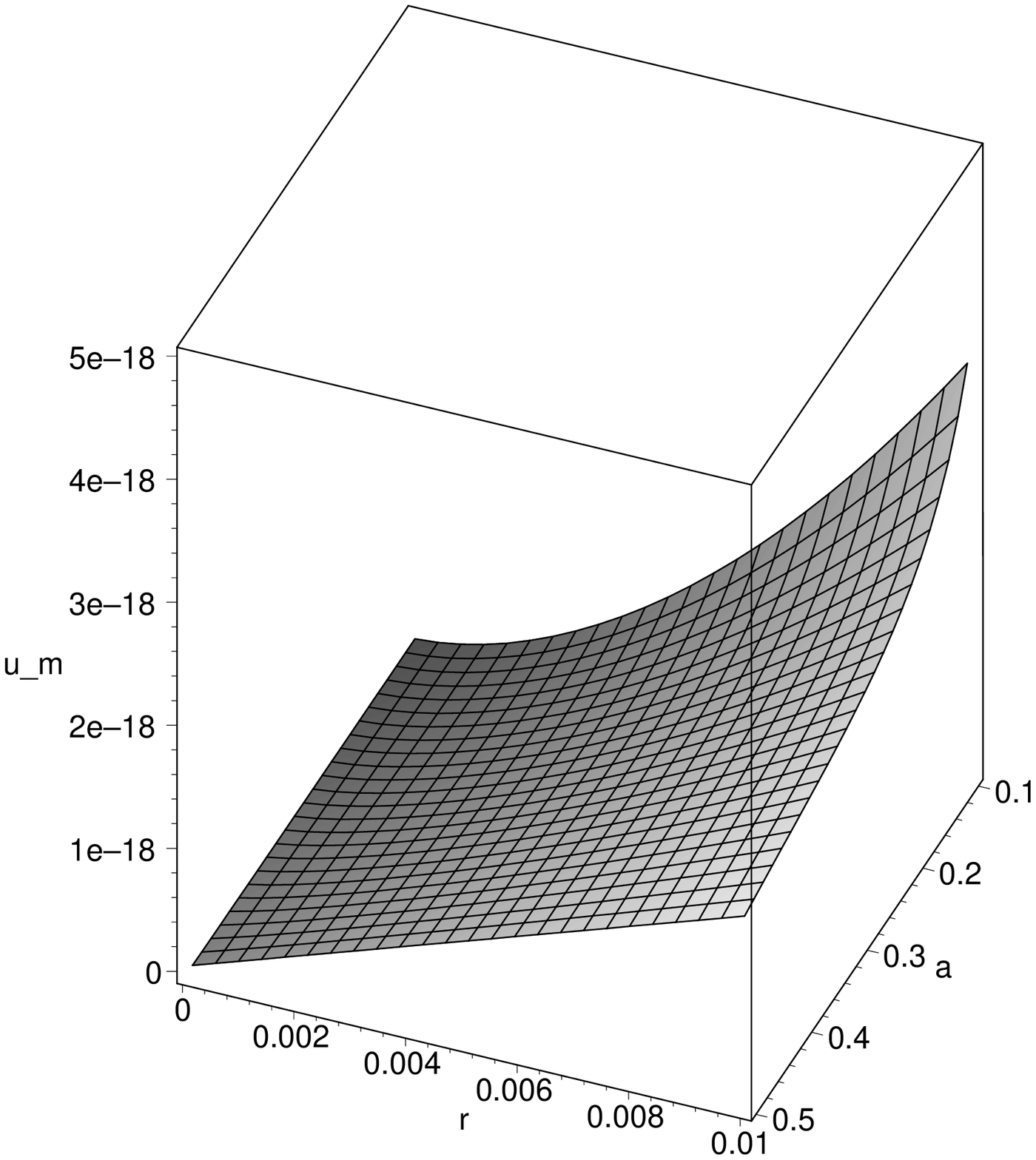}}}&
\rotatebox{0}{\scalebox{.5} 
{\includegraphics{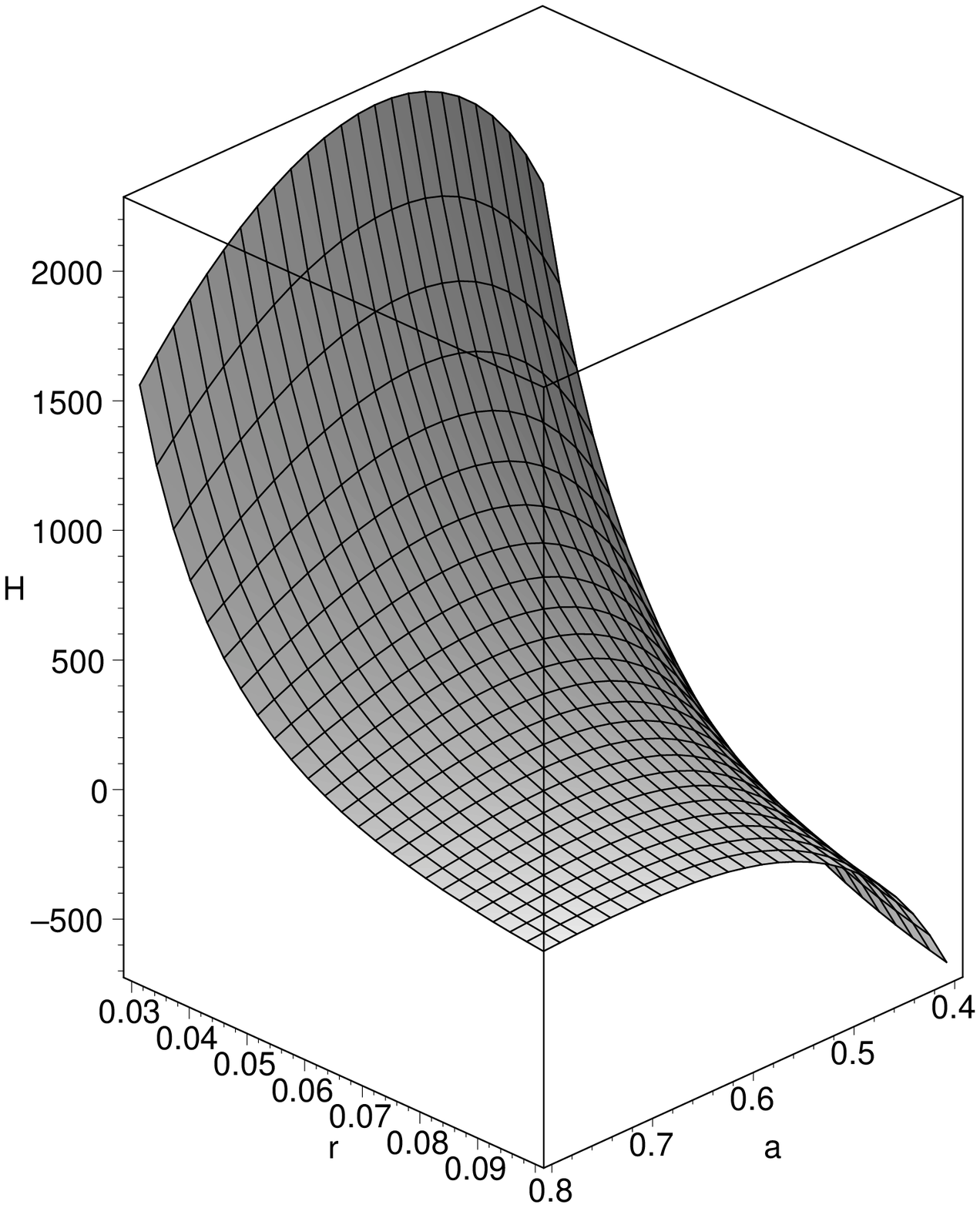}}}\\
\end{tabular}
\caption{The left panel shows the surface $u_m(a,r)$ over our domain of interest with $k=250$ and $E_m=10^{17}$. The right panel shows for the same parameters the surface of constant Boltzmann H function to be compared to the corresponding plot in figure (\ref{fig:H+F}).  }
\label{fig:prelim}
\end{figure}

As a check on this assumption we plot in figure (\ref{fig:prelim}) $u_m(a,r)$ over the domain of interest for our reference parameters. We see that it is approximately constant at the value $10^{-18}$ that we use below. The right panel shows the surface of the Boltzmann H function with variable $u_m$ to be compared to the corresponding plot in figure (\ref{fig:H+F}) for constant $u_m$. We see that, but for the the unphysical region at large $a$ and small $r$, the correspondence is good. We use the Boltzmann function in our analytic arguments of the next section. 

As we discuss somewhat further in the next section, both of the parameters $k$, $u_m$ are important to the behaviour of the density and pseudo-density. This implies a dependence on the central density of the system ($k$) and on the rate at which high energy particles are absent from central regions ($u_m$). There is then strictly a lack of universality in the predicted profiles. However the dependence on the central density could be scaled away at least at small $a$, and the dependence on $u_m$ is relatively weak except at values much closer to unity. 

We have thus chosen a particular physical system as our reference, which has  parameter values $k=250$ and $u_m=10^{-18}$. Such a system (see results below) yield plausible radial profiles (by which we mean similar to those found in simulations) both for $\rho(r)$ and  the pseudo-density $\phi(r)$ out to the radial limits of our assumptions and close to the inner limits of the simulations. Our outer limits are set by the efficiency of the relaxation and to extend them will require adding in the higher order terms in equation (\ref{result1}).  

 The `pseudo phase-space density' $\phi$ (Taylor and Navarro, 2001; Hansen, 2004) may be found simply from equations (\ref{Xdensity}) and (\ref{rms}) as (strictly $e=1$ for their quantity) 
\be
\phi\equiv \rho/\sigma^{3e}.\label{pseudo}
\ee
The explicit dependence on radius is $\propto r^n$ (e.g. paper I and Hansen (2004))
\be
n\equiv (3e-2)a-3e.
\ee
  As for the density the variation of $\phi$ with $r$ is no longer simply given by the power $n$ because of the locus $a(r)$. 

However from the analytical treatment one would like to deduce the variation of $\beta$ or even $a$, at least over the limited range in $r$ wherein the fitting formula is verified. To this end we examine the Boltzmann H function (an entropy density) that follows from our zeroth order equilibrium. We might expect the microphysics of dark-matter halos to produce relative constancy of this function along the locus $a(r)$, since $a$ is monotonic in time at each $r$ until equilibrium is reached\footnote{A recent paper (salvador-Sole et al., 2007) suggests that essentially equation (\ref{asim}) with the appropriate history of $n$ is sufficent to determine $a(r)$. This would oppose the notion of universal central relaxation used here, although the two assumptions appear to yield compatible results in the core.}.

\subsection{H Function and the Locus $a(r)$}

Although it is  understood that a self-gravitating system may not have a condition of maximum entropy globally since this leads to an isothermal DF and an infinite mass, it is possible that locally  (e.g. in the central regions)  a dark matter halo may maximize an entropy density. This leads us to consider the stationary behaviour of the Boltzmann H function that follows from the zeroth order of (\ref{result1}). This is formally
\be
H_o\equiv \int~f_o\ln{f_o}~d^3v.
\ee
Proceeding with a straightforward but tedious calculation of the integrals we obtain explicitly 
\bea
&H_o(a,r;u_m)=\\ 
&4\sqrt{2}C_{oo}~\Phi_o^{(\frac{a}{a-1})}~\left((\ln{\frac{C_{oo}}{\pi}}+\frac{(3-a)}{2(a-1)}~\ln{\Phi_o})D(a;u_m)+\frac{(3-a)(1-a)}{2}\partial_a~D(a;u_m)\right).\nonumber\label{H}
\eea 

\begin{figure}[ht!]
\begin{tabular}{cc} 
\rotatebox{0}{\scalebox{.5} 
{\includegraphics{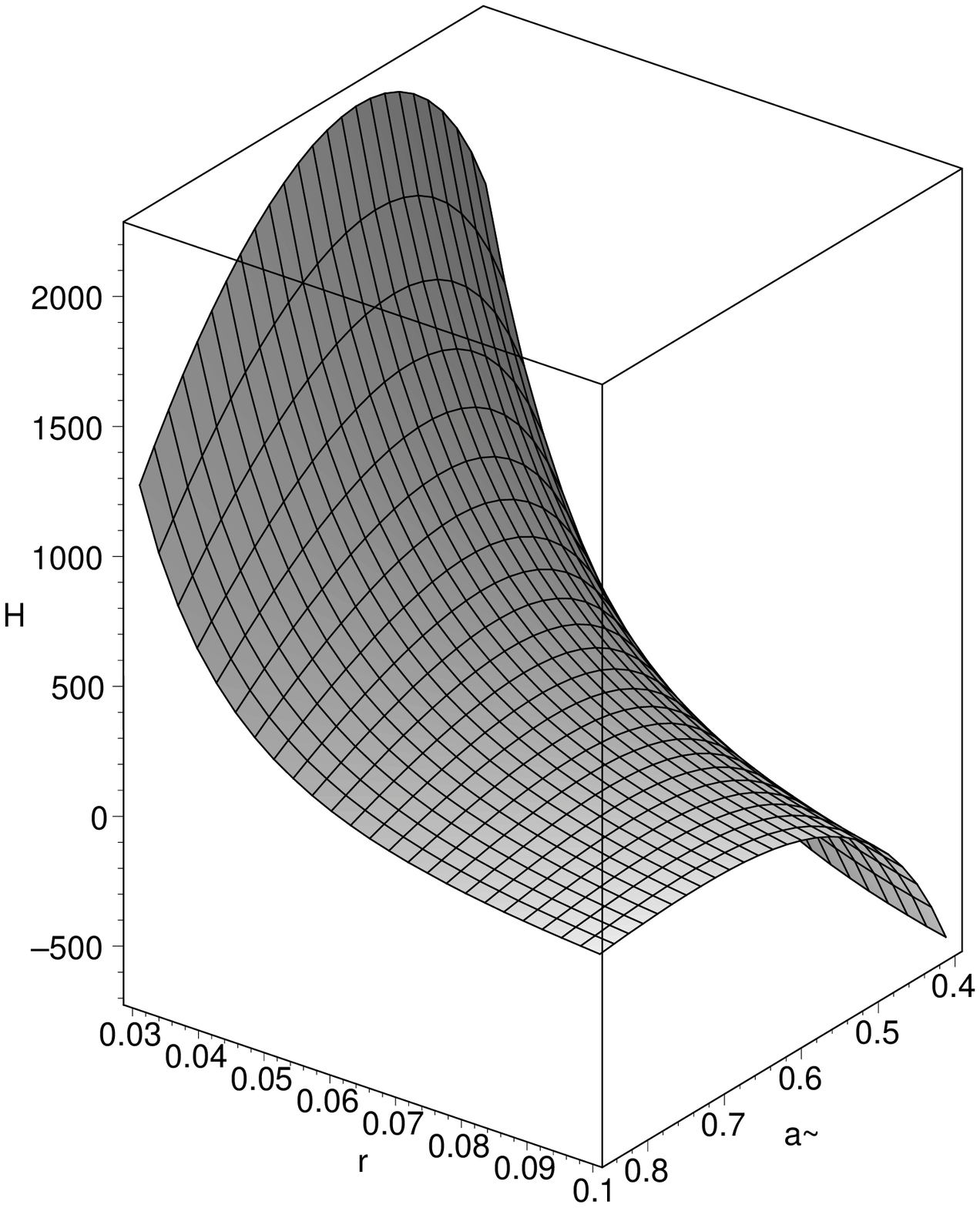}}}&
\rotatebox{0}{\scalebox{.5} 
{\includegraphics{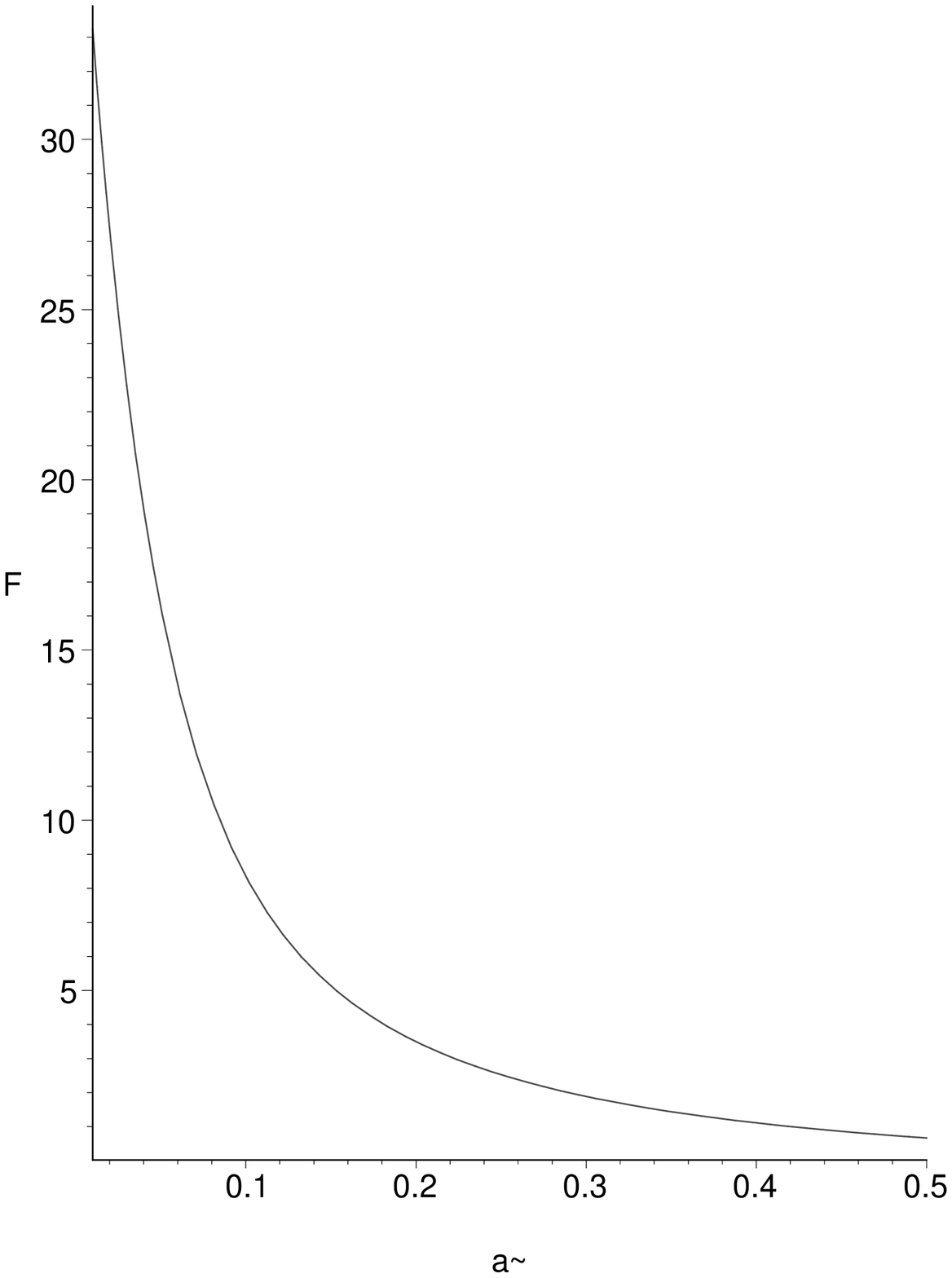}}}\\
\rotatebox{0}{\scalebox{0.5}
{\includegraphics{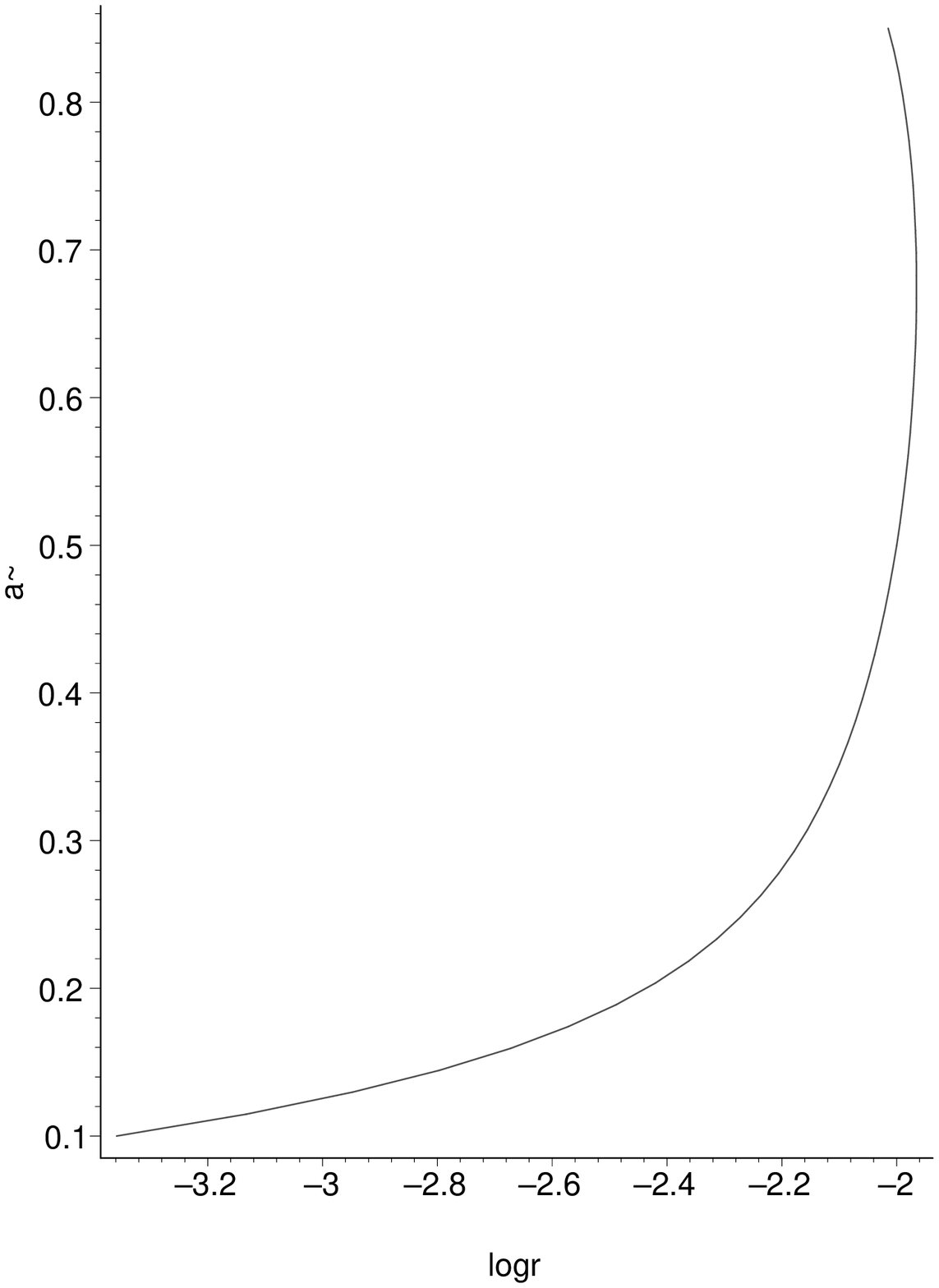}}}&\\
\end{tabular}
\caption{On the left we show part of the surface of the Boltzmann H function over $a$ and $r$ for $k=250$ and $u_m=10^{-18}$. The locus of interest is given by $\delta H(a,r)=0$. The right-hand panel shows $F(a,u_m)\equiv D(a,u_m)$ going to a finite limit as $a\rightarrow 0$. The left bottom panel shows the locus $a(r)$ of stationary values of $H_o$ that follows from equation (\ref{dlocus}) in an interesting range of $a$. We used $k=250$, $u_m=10^{-18}$, and $r=.01$ at $a=0.5$ in the calculation.}
\label{fig:H+F}
\end{figure}

This surface is shown in figure (\ref{fig:H+F}) over an interesting range in $a$, and  $r$. There is in addition an arbitrary fiducial unit for $r$.
The parameters $u_m=10^{-18}$ and $k\equiv 2\sqrt{2}C_{oo}=250$ as noted above. We begin to see the divergence at small $a$ that exists strictly for zero $u_m$. The accompanying graph in figure (\ref{fig:H+F}) shows the behaviour of the function $F$ for the same $u_m$ (independent of $k$). The limiting value as $a\rightarrow 0$ is $\approx 40.833$

We observe that, by using the equilibrium distribution function that is the zeroth order of result (\ref{result1}), we calculate an $H_o$ that is approximately constant compared to the value at large $a$ and small $r$. This latter case does not arise physically. The behaviour is what one might expect in an equilibrium state.
Our additional hypothesis is that we may deduce the variation $a(r)$ by seeking the precise  stationary locus $\delta H_o(a,r)=0$, for which we must solve 
\be
\frac{dr}{da}= -\frac{\partial_a H_o}{\partial_r H_o}.
\label{dlocus}
\ee

 The solution to this differential equation is shown in figure (\ref{fig:H+F})
for $u_m=10^{-18}$ in the left bottom panel. In this example we start at $r=.01$ and $a=0.5$. Some variation of the starting values is possible. We observe that there is a natural outer limit to the usefulness of our procedure where $a$ becomes double-valued. We regard this point as the outer limit to the necessary relaxation, since lower densities (proportional to $k$) causes this limit to move inward. This particular curve is illustrative only, and some fine tuning in $k$ (proportional to the central density) and starting value would be necessary for any particular halo. The physical scale depends on the fiducial unit, but the relative scale is fixed.
However this theoretical curve does indicate that  $a$ varies slowly with $r$, as is suggested by the  numerical simulations (e.g. Navarro et al., 2004).

\begin{figure}[ht!]
\begin{tabular}{cc} 
\rotatebox{0}{\scalebox{.5} 
{\includegraphics{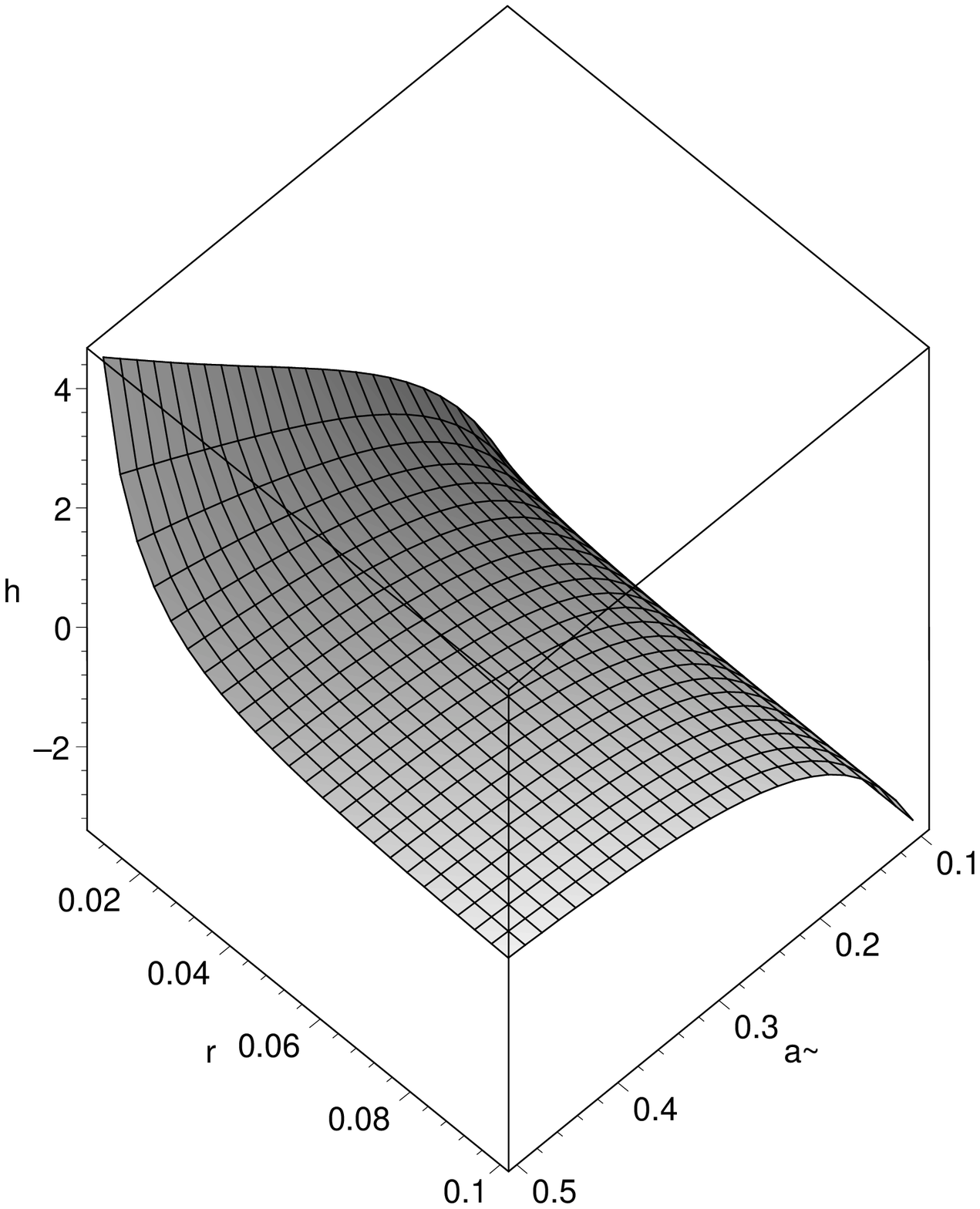}}}&
\rotatebox{0}{\scalebox{.5} 
{\includegraphics{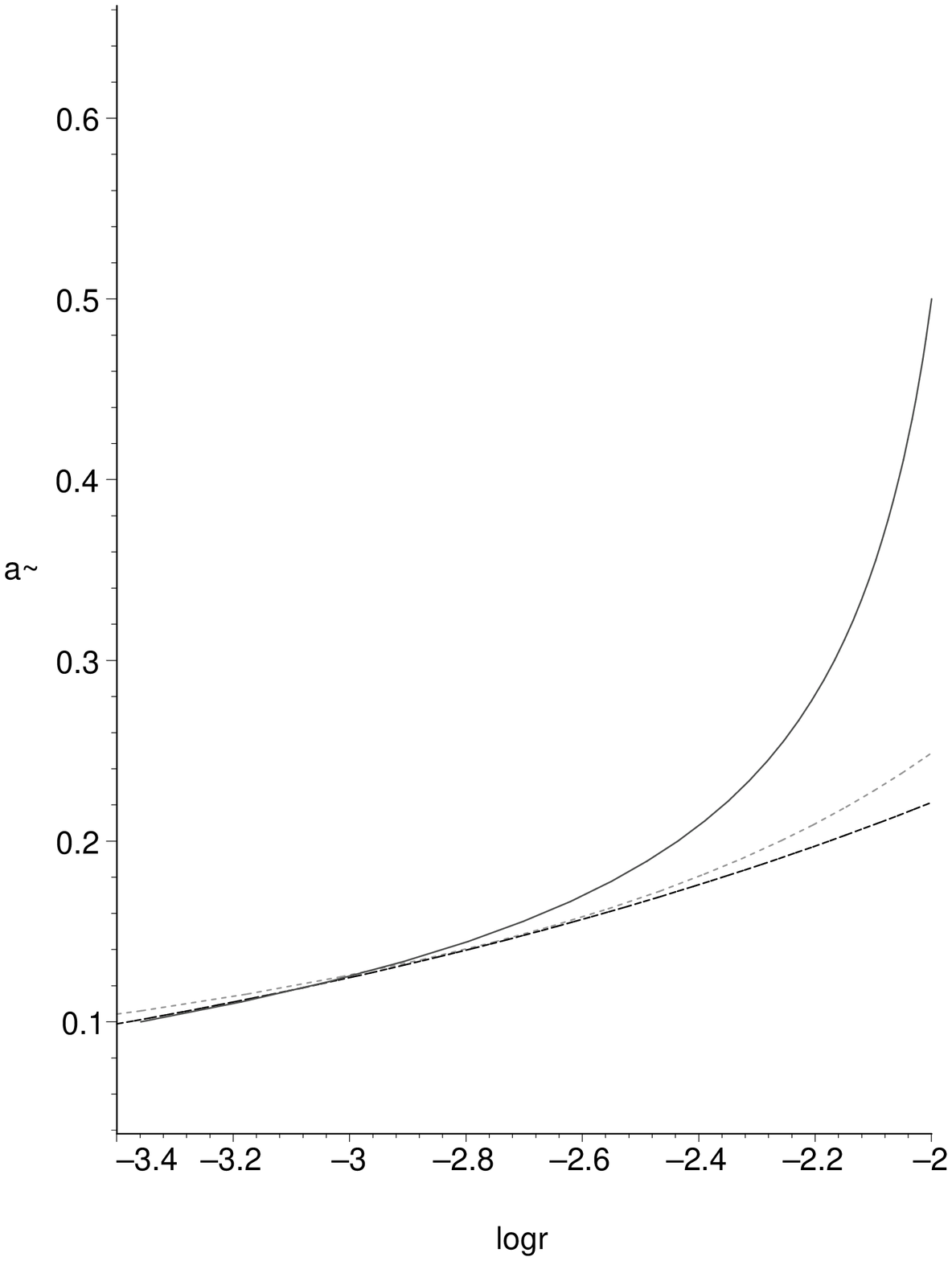}}}\\
\rotatebox{0}{\scalebox{0.5}
{\includegraphics{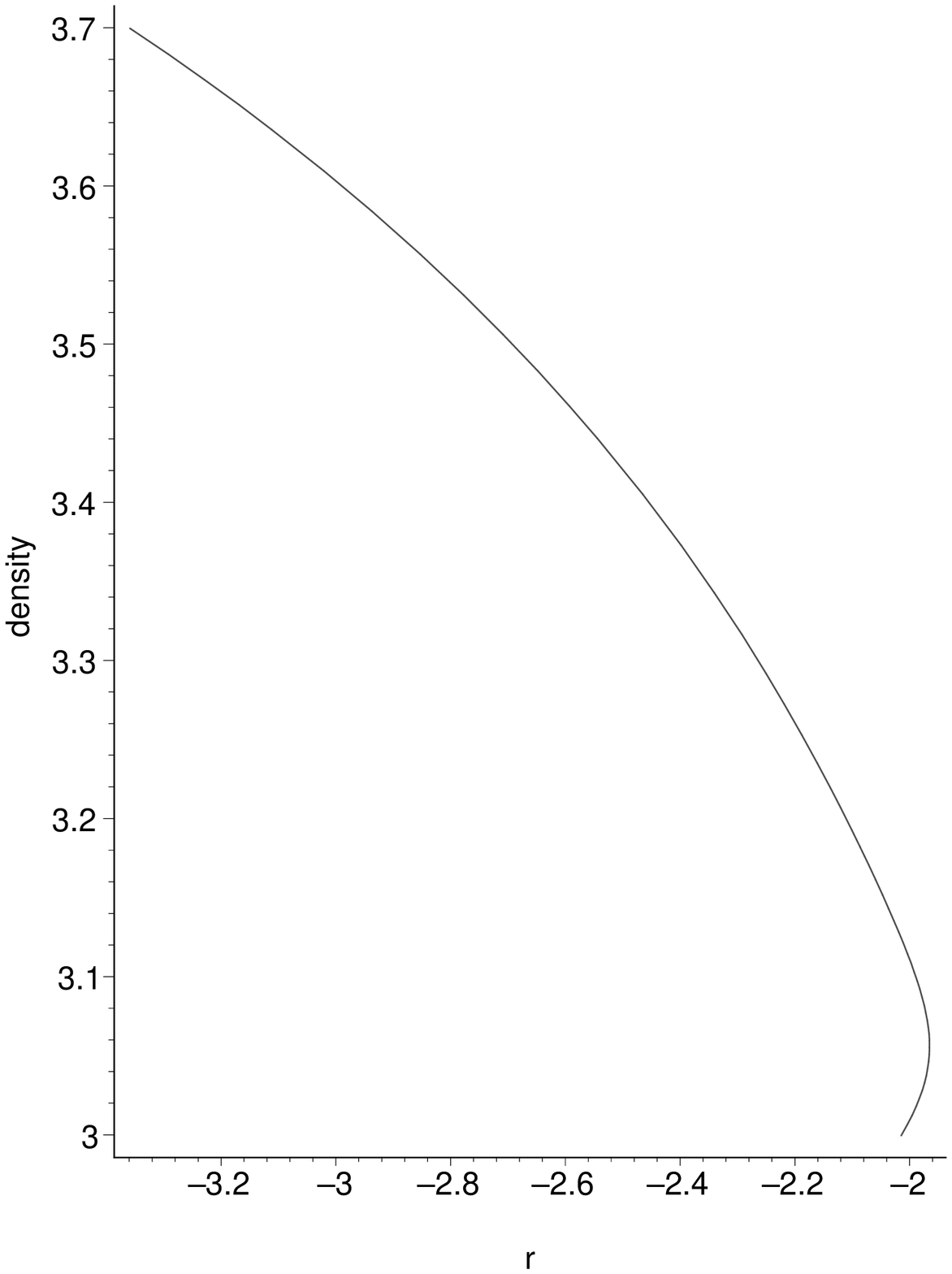}}}&
\rotatebox{0}{\scalebox{0.5}
{\includegraphics{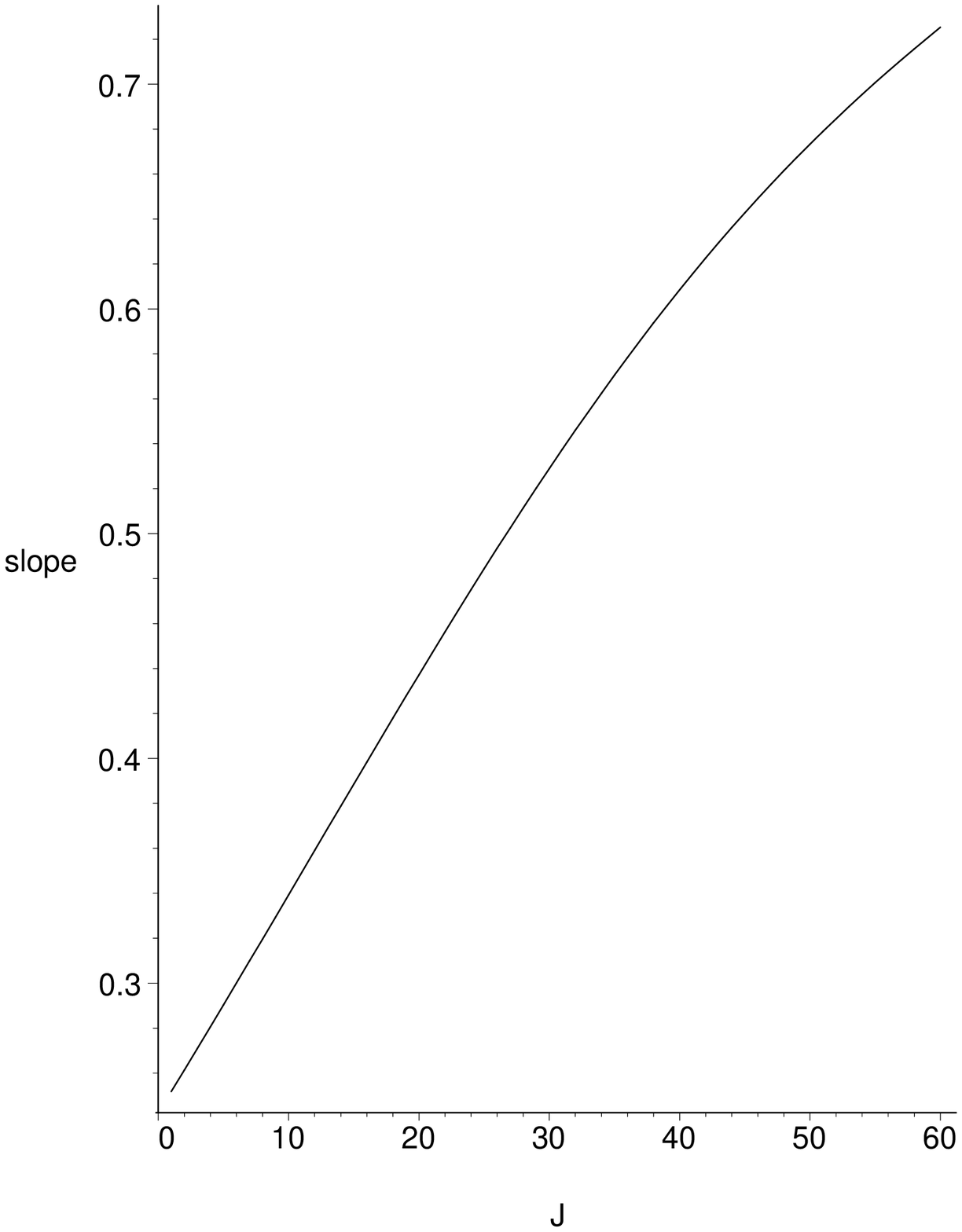}}}\\
\end{tabular}
\caption{The left upper panel shows the suface $h\equiv aH/k$ for $k=250$ and $u_m=10^{-18}$.  Taking the approximation of equation (\ref{locus}) with $h=-1.0$ one obtains the dash-dotted line on the right-hand panel, while the numerical solution (started again at $a=0.5$ with $r=0.01$) is shown as the solid line. The dashed line on this panel is $a=0.7r^{0.25}$. The lower left panel shows the actual density for this case  calculated from equation(\ref{Xdensity}) as a function of log r.  The lower right panel estimates the logarithmic slope $\beta$ as a function of $J$, where $J\approx 247.5(a-0.1)$.  }
\label{fig:h+fits}
\end{figure}

The question now arises as to whether there is an approximate analytic representation of this theoretical locus. Since at small $u_m$  there is a considerable range of $a$ over which the Beta function term in $D(a;u_m)$ dominates; we can hope to find an analytic expression for the locus in the small $a$ region by letting $a\rightarrow 0$ and holding $H_o$ constant, while retaining only this term in $D$. Proceeding in this fashion, but keeping only the largest terms ( including large constants; as well as large constants multiplied by $a$, which is after all finite in the range of interest), we obtain an estimate of the locus $r(a)$ as 
\be
 r=\sqrt{a}\exp{(-1/(2a))}C(a),\label{locus}
\ee
where the variable constant $C(a)$ is actually
\be
C=(\frac{3}{2k})^{1/6}(\frac{3}{2\pi})^{1/3}\exp{-(\frac{H_o}{6k}a)}.
\ee
 The first two factors in $C$ are weakly dependent on $k$ so that the constant depends on $aH_o/(6k)\equiv h$ as $a\rightarrow 0$ but finite. We show this variation in the left upper panel of figure (\ref{fig:h+fits}) for $k=250$ and we observe that but for the region of large $a$ and small $r$ which is unimportant physically, it is relatively constant. The right panel shows the resultant illustrative fit ($h=-1.0$) to the numerical locus (solid line) as a dash-dotted line over the region where we believe our arguments to be justified. A locus of the type suggested by the simulations (Navarro et al. 2004) (although outside their stated parameter range by about two sigma) is shown as the dashed line. The approximation diverges from the exact locus of the stationary H function (solid line) where  expected, but interestingly it continues to fit the simulation-type curve over a wider range in $r$. 

If  one solves equation (\ref{locus}) allowing the exponential factor to vary, one finds two branches. One of which (the smaller $r$ at a given $a$) is the branch that matches the other estimates, and the other is clearly unrealistic.

In the bottom row of  figure (\ref{fig:h+fits}) we show on the left panel the density from equation (\ref{Xdensity}). The outer limit to the assumption of equilibrium is clear where it becomes double valued. The slope is shown on the right hand lower panel to $a\approx .35$. At $a=0.4$ and at $a=0.5$ the slope is $0.794$ and $0.98$ respectively. Thus we see that in this interesting range $2a$ is in fact a reasonable estimate of $\beta$. Moreover the slope begins near the  value $1$, which occurs frequently in the simulations. However it rolls over to a flatter value rather faster at first than seems to be indicated by the simulations.  

We should remember in considering these figures that there are arbitrary units, which must presumably be chosen so as to fit smoothly to the outer well-established NFW behaviour. The dynamical variation of $a$ in this region remains mysterious, although it must have something to do with the evolution of the radial orbit instability (see discussion).

This rough continuity with the simulations does not hold universally however. For lower central density the results are dramatic. For $k=100$ the logarithmic slope never gets larger than $\approx 0.38$ at $a\approx 0.3$, and $2a$ does not approximate $\beta$ well until $a\le 0.2$. Hence for lower densities the domain of applicability of this argument from relaxation moves distinctly to smaller radii and smaller $\beta$. We might therefore conclude that the central density must be sufficiently high for the relaxation to overlap substantially with decreasing density. This seems intuitively reasonable, but does not seem to be sufficient to guarantee the simultaneous presence of the power-law behaviour in the pseudo-density,$\phi$ .

\begin{figure}[ht!]
\begin{tabular}{cc} 
\rotatebox{0}{\scalebox{.5} 
{\includegraphics{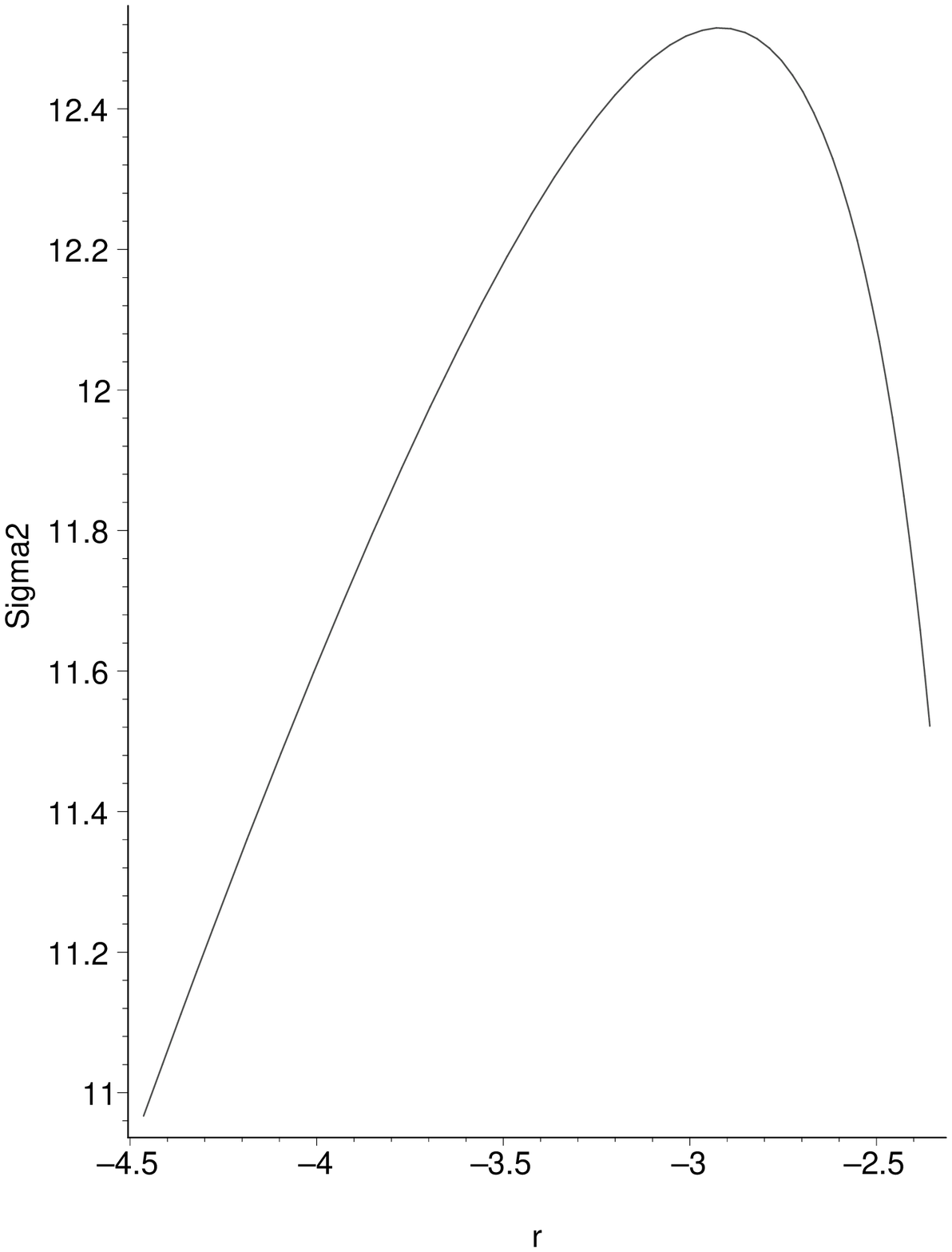}}}&
\rotatebox{0}{\scalebox{.5} 
{\includegraphics{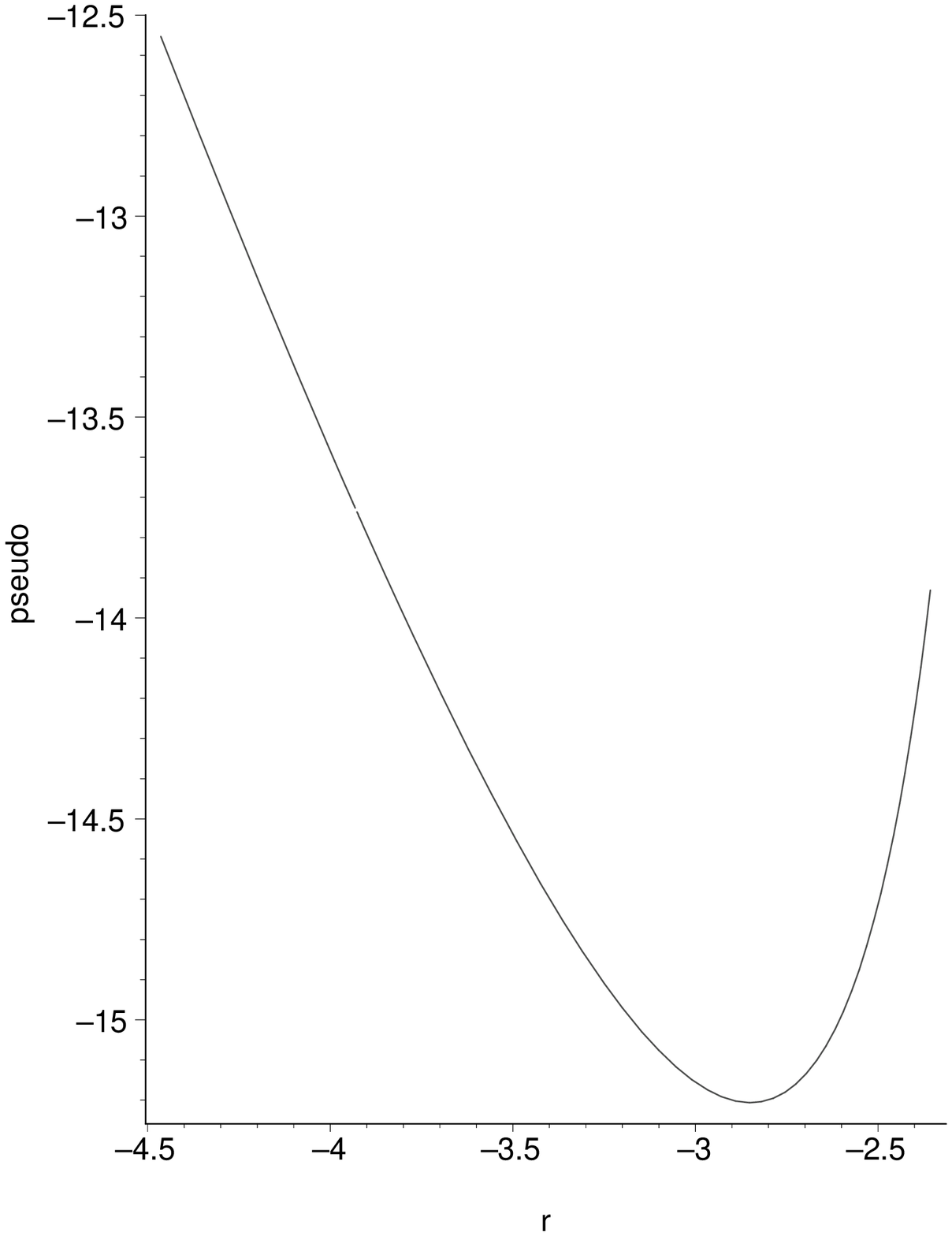}}}\\
\rotatebox{0}{\scalebox{0.5}
{\includegraphics{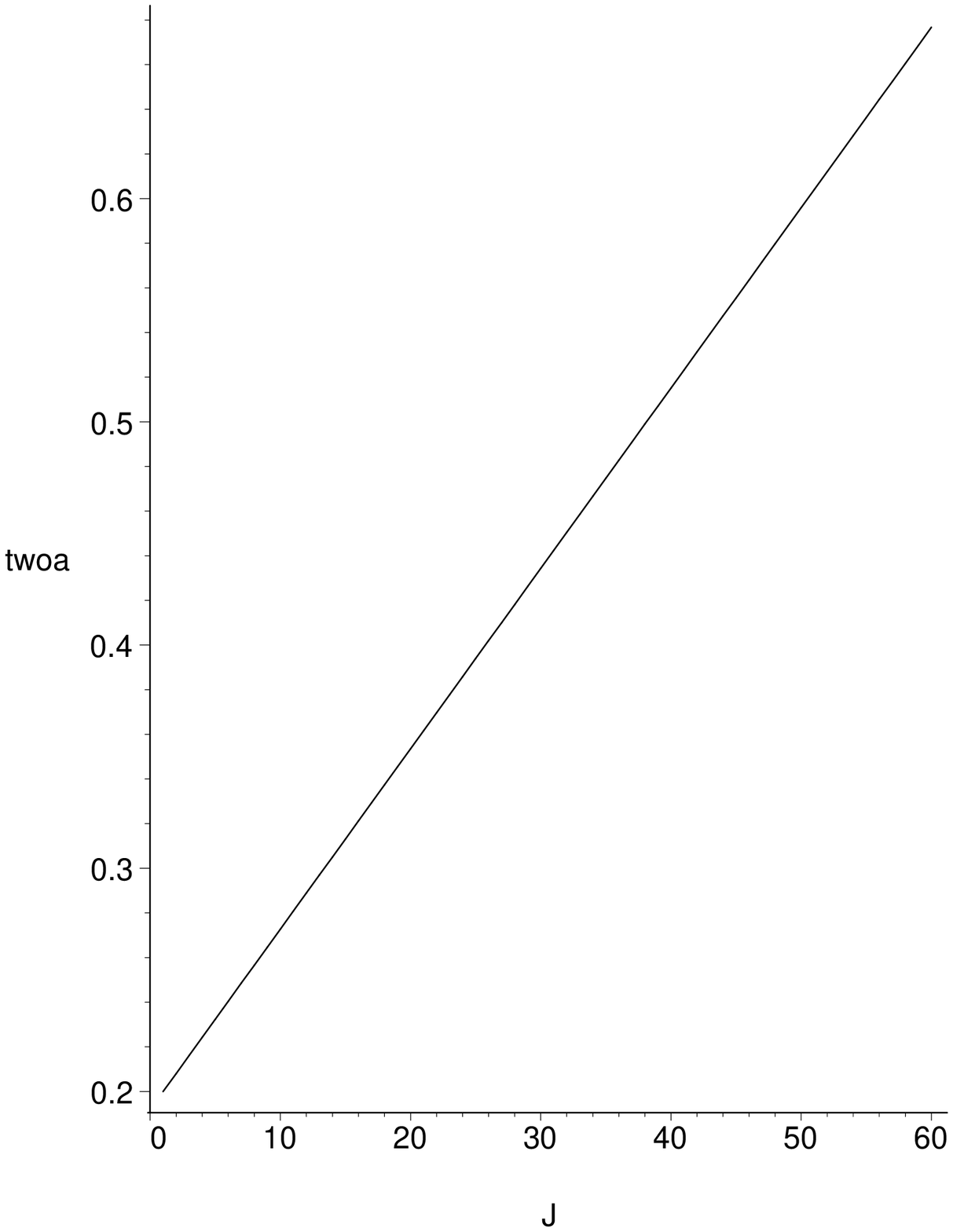}}}&
\rotatebox{0}{\scalebox{0.5}
{\includegraphics{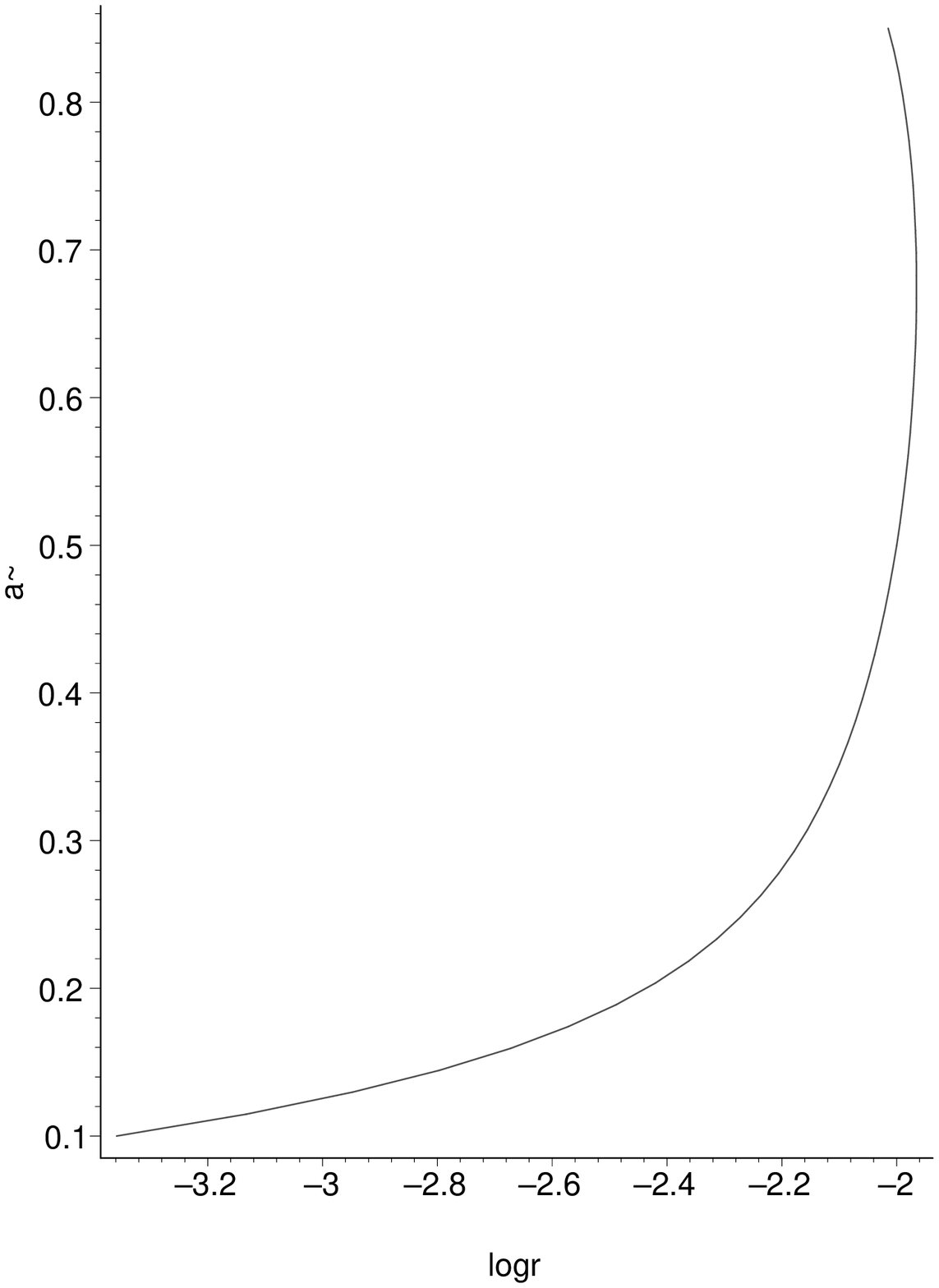}}}\\
\end{tabular}
\caption{The left upper panel shows the variation in $\sigma^2$ with $r$ on a log-log scale while the upper right panel shows the `pseudo-density' on a similar scale. 
The lower left panel plots the linear relation between the naively predicted logarithmic slope $2a$ and the index $J$. The lower right panel repeats the solution to equation (\ref{dlocus}) for convenience. All of these graphs are for our reference example with $k=250$ and $u_m=10^{-18}$.  }
\label{fig:mnemonic}
\end{figure}

This  puzzle arises as we consider both the isotropic velocity dispersion $\sigma^2$ and `pseudo phase-space density' $\phi$. Plots of these quantities are shown in our final figure (\ref{fig:mnemonic}) of this section, together with the linear relation between $J$ and $a$ and the previous calculation of $a$ versus $log (r)$. This latter pair is simply a convenient device for relating $J$ to $a$, $\beta$, and the spatial scale $r$.

We see that the desirable power law in the pseudo-density exists only out to $r\approx 10^{-3}$ where the velocity dispersion peaks. Inside this radius the density is rather flat so that it seems that a flat core must be forming before the pseudo-density becomes a power law contrary to what is observed in the simulations.

This behaviour is characteristic over a broad range of parameter space ($k$ and $u_m$). For example at $k=150$ the pseudo-density power-law has moved out only to $\approx 10^{-2.5}$ where the density is already quite flat. By taking $k=10$ and $u_m=10^{-7}$ one can move the pseudo-density powerlaw out towards continuity with the NFW regime, but now the density is flat and rising slightly (not really a power law but a typical logarithmic slope is $-0.2$). This may be an effect of holding $u_m$ constant at too large an $r$, since effectively this cuts off the high energy particles( We do expect this to happen  as $r\rightarrow 0$ to ensure a finite density). Indeed one finds that holding $E_m$ constant leads in this case to a substantial variation (factor $4$) in $u_m$ in our domain of interest at finite $r$.

Clearly both $k$ and $u_m$ are significant parameters and more detailed study is required. Such a study would make the present work far too long. It probably will not be solved by including the first order term in (\ref{result1}), since this was done in paper I with similar results. For the moment it seems that either the density in the simulations should flatten rapidly in the next order of magnitude in decreasing $r$, or the power-law behaviour of the pseudo-density should break in the same region; if our hypothesis is correct and/or the measures from the simulations are consistent. 

We conclude from this preliminary study that at least qualitatively the stationarity of the H function seems in accord with the empirical behaviour found in the simulations at the inner limit. However the details are discordant in general and there is  a lack of universality at different central densities, although this is mainly a question of changing the scale at which similar behaviour appears.   The microphysics of the inferred relaxation remains somewhat mysterious, although we believe it to be intimately related to a cascade from the radial orbit instability. We discuss these points further in the concluding section

\eject
\section{Discussion and Conclusions}

A key technical result of this paper is a successful renormalization in $u$ of the series that underlies the basic calculation of paper I in section 4.2. This reassures us that the consequences of using a divergent series to study the pseudo-density has not  led us astray. For the critical quantity termed `ratio' in paper I is, asymptotically in time, independent of the renormalization factor (i.e. $1/\kappa_2$). The conclusions drawn there are thus qualitatively unchanged by the renormalization. Even though in that paper we had no natural way of calculating $a(r)$ the results at a given $a$ are rather similar to those found here. 

 We have also confirmed the asymptotically time-independent form (zeroth order DF) for the distribution function at fixed $r$. This is defined to within arbitrary functions of $\kappa_1$ in principle, but the requirement of velocity space isotropy renders these as true constants. The development of isotropy requires a physical mechanism for expanding the phase space available to the system. This we attribute to the radial  orbit instability (see e.g. MacMillan, Widrow and Henriksen, 2006). This equilibrium distribution function is stable according to the usual theorems (e.g. Binney and Tremaine, 1987). 

There is a first order correction to the strict equilibrium DF which is now renormalized in time.  This permits an estimate of the size of the relaxed core ($a\ne 0$) by taking it equal to the zeroth order. The estimate is energy dependent being roughly  $r_{core}\propto |E|^{1/(2(1-a))}$, although we should remember the adiabatic dependence of $a$ on $r$. This result indicates that at any $r$ it is the high energy particles (recall that $E>0$ for the equilibrium result) that are the more likely to be relaxed. This is consistent with such particles having large turn-round radii even if they are relaxed near the centre of the system.  
 
Our second result is the deduction from equation (\ref{dlocus}) of a form for the variation of the logarithmic slope of the density (related to $a(r)$) that is not qualitatively different from that found empirically (Navarro et al., 2004). The expected qualitative behaviour with $r$ is also inferred for the velocity dispersion and the pseudo phase-space density. 

However there is a slight contradiction between this latter behaviour and that of the density. The pseudo-density is more like the simulations at a given central density for  lower energy particles at the centre (large $u_m$), while the density profile is not so dependent on this parameter. There is some evidence that there may be an optimum value for matching the density profile to the simulations however ($u_m=10^{-10}$ is slightly better than larger and smaller values), which is in accord with the lower energy particles being more relaxed.

We have concluded from these considerations that, assuming correctness on both the theoretical and simulative sides, either the density profile or the pseudo-density profile should change dramatically on proceeding to a scale smaller by an order of magnitude than accessible currently. 

If the disagreement with the simulations persists as strongly as at present, we might suspect that something is missing in the theoretical formulation. It is possible that some form of collision term is necessary in the Boltzmann equation in order to describe properly the wave-particle interaction.

 We have used the stationarity of the Boltzmann H function to deduce these profiles. However this argument does not by itself provide a physical mechanism for the onset of such a condition. We must identify some scattering mechanism that increases the phase-space volume accessible to the system, relative to that available to nearly radial infall. 

The likely candidate appears to be the radial orbit instability (e.g. Barnes et al. 2005, 2006; Lu et al.,2006; MacMillan, Widrow and Henriksen, 2006) plus a subsequent cascade to smaller scales. Such a cascade might be due to the wave-particle interaction as discussed in Henriksen (2006a) and elaborated below. The self-similar form of the squared angular momentum that follows from equation (\ref{vars}) namely $j^2=\zeta_2^2 r^{(4-2a)}$ already requires something like the radial orbit instability to generate this quantity. Moreover we see from equation (\ref{acceleration}) below that $j$ is proportional to $\sqrt{Gmr}$, which behaviour has been associated with the radial orbit instability and indeed previously with the existence of self-similarity (MacMillan, Widrow and Henriksen, 2006 and references therein). 

The process is presumably working during the evolution from the $r^{-2}$ density profile to that of $r^{-1}$ and flatter. However our argument from `thermodynamic' equilibrium does not extend to this region. The evolution must then be occurring without the local maximization of entropy such as represented by a cascade.

 Since the locus $a(r)$ presumably contains  the relaxation physics, it is of interest to see how it appears in various dominant physical quantities during the approach  to the `relaxed' region from the outside.  
 
Let us first recall that the mass $m(r,t)$ of the growing halo is given by (see e.g. Henriksen and Widrow, 1999 ) 
\be
m(r,t)=M(R)~e^{((3/a-2)\alpha T)},
\ee
which on recalling equation (\ref{vars}) becomes 
\be
m(r,t)=M(R)(r/R)^{(3-2a)}.
\ee
Consequently we obtain 
\be
\frac{Gm}{r^2}=\frac{GM}{R^{3-2a}}r^{(1-2a)}.\label{acceleration}
\ee
 
The first factor on the right-hand side of this last result is constant at a fixed fraction of the growing halo (i.e. constant $R$), so that the  physical gravitational acceleration acting on particles there varies as $r^{1-2a}$.

Starting at $a=1$  which applies  outside the isotropic core where the velocity is predominantly radial  and lies outside our calculation from the Boltzmann H function, we  can describe  a systematic decrease in the dependence of the radial acceleration on $r$ as $a$ decreases.  This should correspond to decreasing susceptibility to the radial orbit instability (see e.g. MacMillan, Widrow and Henriksen, 2006). In fact however it is probably the action of the radial orbit instability that decreases $a$ initially. We recall that by the approximate self-similarity we expect $\beta\approx 2a$ in this regime with $a$ locally constant.  

At $a=1$ the radial gravitational acceleration at a constant fraction of the growing core varies as $r^{-1}=R^{-1}(\alpha t)^{-1}$. Hence it is very strong initially ($t\rightarrow 0$) at all $R$ and creates predominantly radial infall with a density profile $\rho\propto r^{-2}$, a constant $\sigma$, and hence a pseudodensity power $n=-2$. This state is highly susceptible to the radial orbit instability (MacMillan, Widrow and Henriksen, 2006). We ignore the $r^{-3}$ outer cut-off here, which has been described elsewhere as due either to mass exhaustion or tidal truncation ($r^{-4}$ in that case: e.g. Henriksen, 2004). 

We suppose $a$ to decrease with $r$ as a consequence of the instability until it reaches $3/4$, which is a limiting value found by Moore et al. (1999). This yields a weaker radial acceleration  proportional to $r^{-1/2}=R^{-1/2} (\alpha t)^{-2/3}$.  At any fixed core fraction we should  expect this phase at a sufficiently late time, but it arises at small core fractions first. The weaker radial acceleration presumably coincides with the trend towards isotropy in the core, since the radial orbit instability has already acted most efficiently.

 For $a=3/4$ one finds also  $\rho\propto r^{-3/2}$, $\sigma\propto r^{1/4}$ and $n=-(3/2)(1+e/2)$, by treating $a$ as locally constant in the full expressions. This latter value is $-2.25$ for $e=1$. This may seem too large compared to the value of $n$ found in cosmological simulations (e.g. Dehnen and McLaughlin, 2005), but interestingly it corresponds quite closely to the value found by MacMillan (2006) in a study of the collapse of an isolated halo with cosmological initial perturbations. He finds globally $n=-2.19\pm .03$, and a logarithmic density slope close to $\beta=-1.5$ is a reasonable mean value over about a factor 4 in radius. Moreover $\sigma$ increases roughly as $r^{1/4}$ in these core regions according to MacMillan's simulations. It is possible that the steeper value of $n$  is an effect of the isolated nature of this halo compared to the cosmological simulation, perhaps due to a resulting lower velocity dispersion. In our calculations we find that $\phi$ is steeper when $u_m$ is larger, that is when fewer high energy particles are retained.

Should $a$ attain the value $2/3$ by continued relaxation with decreasing $r$, we obtain locally the Evans and Collett (1997) solution,wherein the radial acceleration at a fixed halo fraction is $\propto r^{-1/3}R^{-1/3}(\alpha t)^{-1/2}$. This phase again becomes more important at smaller $R$ and late times.  We expect the trend towards isotropy to be even more strongly developed here. One finds consistently for this phase with locally constant $a$; $\rho\propto r^{-4/3}$, $\sigma\propto r^{1/3}$ and $n=4/3+e$, that is $n=7/3$ for $e=1$. This re-expresses the transitory steepening in $n$ as described in paper I, but for which there is little evidence in the cosmological simulations.  

The intriguing part of this steady solution is that it can be interpreted as solving the full collisional Boltzmann equation (Evans and Collett, 1997). It is unlikely that particle-particle collisions play a r\^ole in the evolution of dark matter halos, but wave-particle collective interactions, equivalently clump- clump and sub-clump-clump interactions, may well do. Indeed there is a Kolmogorov cascade from larger to smaller radii ($\sigma^3/r=const$) in this solution, which implies inter-scale coupling, and fits well with the idea that a cascade from the radial orbit instability is driving the evolution exterior to the relaxed core. Indeed Henriksen (2006a) has suggested that such a cascade is essential to the relaxation of collisionless matter, where the `clumps' on a given scale were essentially identified with Landau-damped waves. Such a mechanism allied with the radial orbit instability at large scales would create the approach to the stationarity of the H function on this view. 

Pursuing this cascade idea slightly further, we note that $\sigma^3/r\propto r^{(2-3a)}$ locally. Thus for $a<2/3$ the energy flux decreases to smaller $r$ and increases to larger $r$ while the converse is true for $a>2/3$. 

This cascade behaviour suggests to us that a state with $a<2/3$ will trap energy, perhaps creating an inverse cascade back to larger radii. This may increase the effective $a$ (thus reducing the radial dependence in $\sigma$) at larger $r$. But if it increases above $a=2/3$, the cascade is now capped at larger radii and it should again reverse. In this way $a=2/3$ may be a stable point in the adiabatic self-similar history of the core, about which slope oscillations may occur. This state is ultimately transitory,
since if entropy is steadily increasing locally through the cascade process, we expect that  considerations such as those given above will apply. This leads to a stationary H function  and the decline in $a$ will continue.
  
In order to obtain the NFW profile we must consider that relaxation continues until $a=1/2$ and beyond . This limit is distinguished by having a constant gravitational acceleration at a fixed fraction of the growing core. It is just at the boundary of what we deem to be the relaxed region so that subsequent behaviour would be as discussed in the previous section.

Overall we have thus explored several new ideas in this work. The technical part uses the renormalization idea pioneered by Kunihiro and McDonald (ibid). We have in addition  examined for the first time the consequences of maximizing entropy locally, given a form for the distribution function that can vary adiabatically. Moreover we have discussed the outer region where this is not likely to be true in terms of an energy cascade from macroscopic instabilities and the strength of the radial acceleration. The instabilities may be internal such as the radial orbit instability, or they may be created by mergers. The results are promising for our understanding, but not yet definitive.


\acknowledgements{
This work was supported in part by the Canadian Natural Science and Engineering Research Council. The author thanks an anonymous referee for a careful reading of the manuscript and constructive criticism. }


\end{document}